\documentclass[sigplan, screen, 10pt]{acmart}

\setcopyright{cc}
\setcctype{by}
\acmDOI{10.1145/3814943.3816174}
\acmYear{2026}
\copyrightyear{2026}
\acmISBN{979-8-4007-2721-4/2026/06}
\acmConference[LCTES '26]{Proceedings of the 27th ACM SIGPLAN/SIGBED International Conference on Languages, Compilers, and Tools for Embedded Systems}{June 15--16, 2026}{Boulder, CO, USA}
\acmBooktitle{Proceedings of the 27th ACM SIGPLAN/SIGBED International Conference on Languages, Compilers, and Tools for Embedded Systems (LCTES '26), June 15--16, 2026, Boulder, CO, USA}

\usepackage{booktabs} 
\usepackage{microtype}

\usepackage{xcolor}
\usepackage{mathtools}
\usepackage{etoolbox}
\usepackage{listings}
\usepackage{graphicx}
\usepackage{multirow}
\usepackage{algorithm}
\usepackage{enumitem}
\usepackage{bm}
\usepackage{float}
\usepackage[htt]{hyphenat}
\usepackage[noend]{algpseudocode}
\algrenewcommand\algorithmicindent{0.5em}
\usepackage{subcaption}
\usepackage{balance}

\algnewcommand\algorithmicswitch{\textbf{switch}}
\algnewcommand\algorithmiccase{\textbf{case}}
\algnewcommand\algorithmicassert{\texttt{assert}}
\algnewcommand\Assert[1]{\State \algorithmicassert(#1)}%

\algdef{SE}[SWITCH]{Switch}{EndSwitch}[1]{\algorithmicswitch\ #1\
\algorithmicdo}{\algorithmicend\ \algorithmicswitch}%
\algdef{SE}[CASE]{Case}{EndCase}[1]{\algorithmiccase\ #1}{\algorithmicend\
\algorithmiccase}%
\algtext*{EndCase}%

\AtBeginEnvironment{algorithmic}{\lineskip0pt}
\makeatletter
\newcommand*{\algrule}[1][\algorithmicindent]{%
    \makebox[#1][l]{%
        \hspace*{.3em}
        \vrule height .7\baselineskip depth .3\baselineskip
    }
}

\newcount\ALG@printindent@tempcnta
\def\ALG@printindent{%
    \ifnum \theALG@nested>0
    \ifx\ALG@text\ALG@x@notext
    \else
    \unskip
    \ALG@printindent@tempcnta=1
    \loop
    \algrule[\csname ALG@ind@\the\ALG@printindent@tempcnta\endcsname]%
    \advance \ALG@printindent@tempcnta 1
    \ifnum \ALG@printindent@tempcnta<\numexpr\theALG@nested+1\relax
    \repeat
    \fi
    \fi
}
\patchcmd{\ALG@doentity}{\noindent\hskip\ALG@tlm}{\ALG@printindent}{}{\errmessage{failed
to patch}}
\patchcmd{\ALG@doentity}{\item[]\nointerlineskip}{}{}{} 
\makeatother

\begin{document}

\title[Memory-Aware Runtime for Adaptive Draft Scheduling in Speculative
Decoding on Edge Devices]{MemSpec: Memory-Aware Runtime for Adaptive Draft
Scheduling in Speculative Decoding on Edge Devices}

\author{Eunjeong Kim}
\orcid{0009-0005-5678-1357}
\affiliation{%
  \institution{Kyungpook National University}
  \city{Daegu}
  \country{Republic of Korea}
}
\email{kimeunjeong23@knu.ac.kr}

\author{Yeong Jun Jeon}
\orcid{0009-0008-7790-9796}
\affiliation{%
  \institution{Kyungpook National University}
  \city{Daegu}
  \country{Republic of Korea}
}
\email{jyj5219@knu.ac.kr}

\author{Myeonggyun Han}
\orcid{0000-0003-1832-1032}
\affiliation{%
  \institution{Kyungpook National University}
  \city{Daegu}
  \country{Republic of Korea}
}
\email{mhan@knu.ac.kr}



\begin{abstract}

Speculative decoding accelerates autoregressive large language model (LLM)
inference by using a lightweight draft model to speculate multiple tokens,
reducing expensive target model decoding steps. Its effectiveness depends
heavily on draft selection, motivating adaptive methods that exploit variation
across inputs and generation stages. On memory-constrained edge devices,
however, these methods often fail to improve end-to-end throughput due to the
overhead of switching between draft models. We identify a key limitation in
this setting: the mismatch between draft selection and draft availability under
tight memory budgets.

To address this challenge, we present \textsc{MemSpec}, a prediction-guided,
memory-aware runtime for adaptive speculative decoding on edge devices.
\textsc{MemSpec} decouples draft selection from execution through proactive
resident working-set management. A lightweight predictor estimates draft
effectiveness from prompt and generation context, while a memory-aware
scheduler reduces reactive model loading overhead. Experiments on a Jetson
Orin Nano show that \textsc{MemSpec} improves steady-state generation
throughput by 40.7\% on average over state-of-the-art bandit-based adaptive
methods while closely approaching the oracle upper bound.

\end{abstract}

\maketitle

\section{Introduction}

Large language models (LLMs) are increasingly moving beyond cloud datacenters
to memory-constrained edge platforms, including mobile devices, embedded
systems, and edge servers~\cite{10906629,10.1145/3649329.3655665}. As on-device
AI adoption accelerates, achieving high-throughput LLM inference under tight
memory budgets has become a critical systems challenge. These constraints
fundamentally reshape the design space of inference optimization techniques.

Speculative decoding~\cite{10.5555/3618408.3619203,10.5555/3692070.3692273,10.1145/3620666.3651335}
has emerged as an effective approach for accelerating autoregressive inference
by using a lightweight draft model to generate multiple candidate tokens, which
are then verified by a larger target model, thereby amortizing expensive
target-model computation. When the draft model accurately predicts the target
model's outputs, speculative decoding can significantly improve throughput
while preserving output quality, making it particularly attractive for
resource-constrained edge deployments.

However, the effectiveness of speculative decoding is highly sensitive to the
choice of draft model. Prior
works~\cite{yi-etal-2024-towards,kim2025a,yan-etal-2025-decoding} show that
different draft models exhibit widely varying token acceptance rates across
prompts and generation stages. In practice, such variation arises naturally
from domain-specialized draft models (e.g., code, math, or legal), which
perform well on their target domains but perform poorly on out-of-domain
workloads. This motivates adaptive approaches that dynamically select or switch
between multiple candidate drafts.

On memory-constrained edge devices, however, this adaptivity introduces a major
systems challenge. Since only a small number of draft models can reside in fast
memory, switching to a non-resident draft incurs substantial loading overhead.
As a result, frequent draft transitions can negate potential throughput gains
and even degrade overall performance.

Existing approaches are not well suited to this setting. Traditional static
methods~\cite{10.5555/3618408.3619203,10.1145/3620666.3651335,liu2025pearl,pmlr-v262-mamou24a}
that use a single fixed draft model fail to exploit generation heterogeneity,
resulting in suboptimal performance. State-of-the-art adaptive methods typically
rely on multi-armed bandit (MAB)-based
approaches~\cite{hou2025banditspec,kim2025a,liu2026notabandit} to explore and
select draft models at runtime, often improving token acceptance rates. However,
these methods incur frequent draft switching and implicitly assume that selected
drafts can be executed immediately. This assumption breaks down on edge devices,
where memory constraints limit draft residency. As a result, the cost of
switching to non-resident drafts often outweighs the benefit of improved draft
selection, and better draft selection does not necessarily translate into higher
generation throughput.

This limitation stems from a fundamental mismatch between \emph{draft
selection} and \emph{draft availability}: a draft that is predicted to be
effective may not be immediately executable under memory constraints. On edge
devices, this mismatch makes draft switching prohibitively expensive, rendering
exploration-based adaptation inefficient. These observations lead to a key
insight: adaptive speculative decoding on edge devices must be formulated not
only as a \emph{selection problem}, but also as a \emph{scheduling problem}
under memory constraints.

To address this challenge, we propose \textsc{MemSpec}, a prediction-guided,
memory-aware runtime for adaptive draft scheduling. Instead of relying on
costly online exploration, \textsc{MemSpec} predicts promising draft models
using both prompt features and recent generation context, and proactively
aligns draft residency with future demand. By decoupling draft selection from
execution, \textsc{MemSpec} always proceeds with the best currently resident
draft while preparing better candidates in the background, thereby avoiding
blocking and excessive model loading overhead.

Specifically, this paper makes the following contributions:
\begin{itemize}[noitemsep,nolistsep]

\item We identify a fundamental limitation of adaptive speculative decoding on
memory-constrained edge devices, namely the mismatch between draft selection
and draft availability. We show that exploration-based methods incur excessive
switching overhead, as selecting a better draft often requires loading
non-resident models, limiting end-to-end throughput gains.

\item We propose \textsc{MemSpec}, a prediction-guided, memory-aware runtime
that decouples draft selection from execution. \textsc{MemSpec} replaces online
exploration with lightweight prediction and enables non-blocking decoding by
always selecting the best currently resident draft.

\item We design a memory-aware scheduling framework that proactively manages a
small working set of draft models under tight memory budgets. Through
coordinated prefetching and eviction, \textsc{MemSpec} aligns draft
residency with predicted future demand and enables non-blocking adaptive
decoding under memory constraints.

\item We implement \textsc{MemSpec} on a Jetson Orin Nano and evaluate it
across diverse workloads. \textsc{MemSpec} improves end-to-end throughput by
58.8\% over static baselines and 40.7\% over state-of-the-art adaptive
methods, while achieving performance close to the dynamic oracle upper bound.

\end{itemize}

\section{Background and Motivation}

This section presents three key observations that motivate \textsc{MemSpec}:
(1) speculative decoding efficiency varies across workloads and generation
stages, creating substantial headroom for adaptive draft selection;
(2) on memory-constrained edge devices, switching to a non-resident draft
incurs high latency, often exceeding several decoding iterations; and
(3) exploration-based adaptive methods repeatedly incur this switching cost,
limiting their effectiveness.

\subsection{Speculative Decoding Efficiency and Acceptance}

In speculative decoding (SD), autoregressive generation proceeds in iterative
draft-and-verify steps. At each iteration, a draft model proposes candidate
tokens, which are then verified by the target model. Let $A$ denote the number
of tokens accepted by the target model in one iteration. For a resident draft,
throughput can be approximated as:
\begin{equation}
\text{Throughput} \approx \frac{\mathbb{E}[A]}{L_{\text{iter}}},
\label{eq:throughput_model}
\end{equation}
where $L_{\text{iter}}$ is the average latency of one SD iteration. Improving
$\mathbb{E}[A]$ is therefore the primary lever for increasing throughput.

We evaluate a quantized \textsc{LLaMA-2} 7B target model (GPTQ INT4) with five
400M-parameter draft models: one general-purpose draft and four
domain-specialized drafts trained for code, mathematical reasoning, legal, and
medical workloads. Detailed setup is described in
Section~\ref{subsec:eval-setup}.

\begin{figure}[t]
\centering
\includegraphics[width=\linewidth]{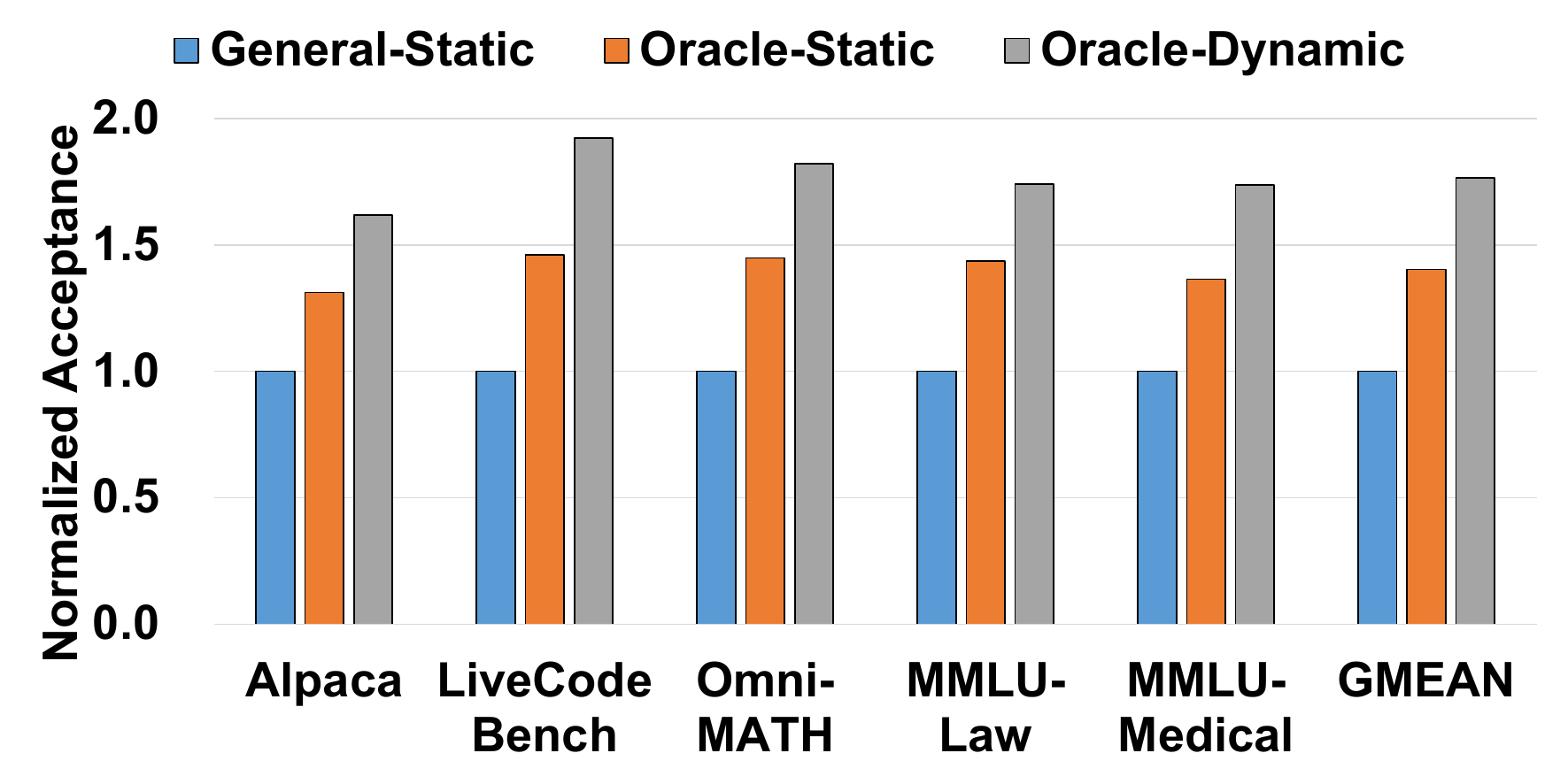}
\caption{Impact of static and dynamic draft selection on token acceptance across datasets.}
\label{fig:acceptance_dataset}
\end{figure}

Figure~\ref{fig:acceptance_dataset} compares three configurations:
\textit{General-Static}, \textit{Oracle-Static}, and \textit{Oracle-Dynamic}.
\textit{General-Static} uses a single general-purpose draft,
\textit{Oracle-Static} selects the best draft per prompt via offline evaluation,
and \textit{Oracle-Dynamic} dynamically switches draft models within a single
generation.

\textbf{Observation 1: Static selection is insufficient due to variability in
draft effectiveness.} Across datasets, \textit{Oracle-Static} improves
normalized acceptance by $40.3\%$ on average over \textit{General-Static},
demonstrating that selecting an appropriate specialized draft model is critical
for performance. Moreover, \textit{Oracle-Dynamic} provides an additional
$25.7\%$ improvement over \textit{Oracle-Static}, indicating that the most
effective draft can change within a single generation. These results show that
even the best static selection is insufficient and that dynamic adaptation is
necessary to fully exploit speculative decoding efficiency.

\subsection{Memory Constraints and Draft Switching Overheads on Edge Platforms}

The headroom identified above does not directly translate into throughput on
edge platforms. Unlike server-class systems, edge devices cannot keep many
draft models resident in fast memory simultaneously. Switching to a better
draft therefore requires loading a non-resident model from a slower memory
tier, such as NVMe storage.

\begin{figure}[t]
\centering
\includegraphics[width=\linewidth]{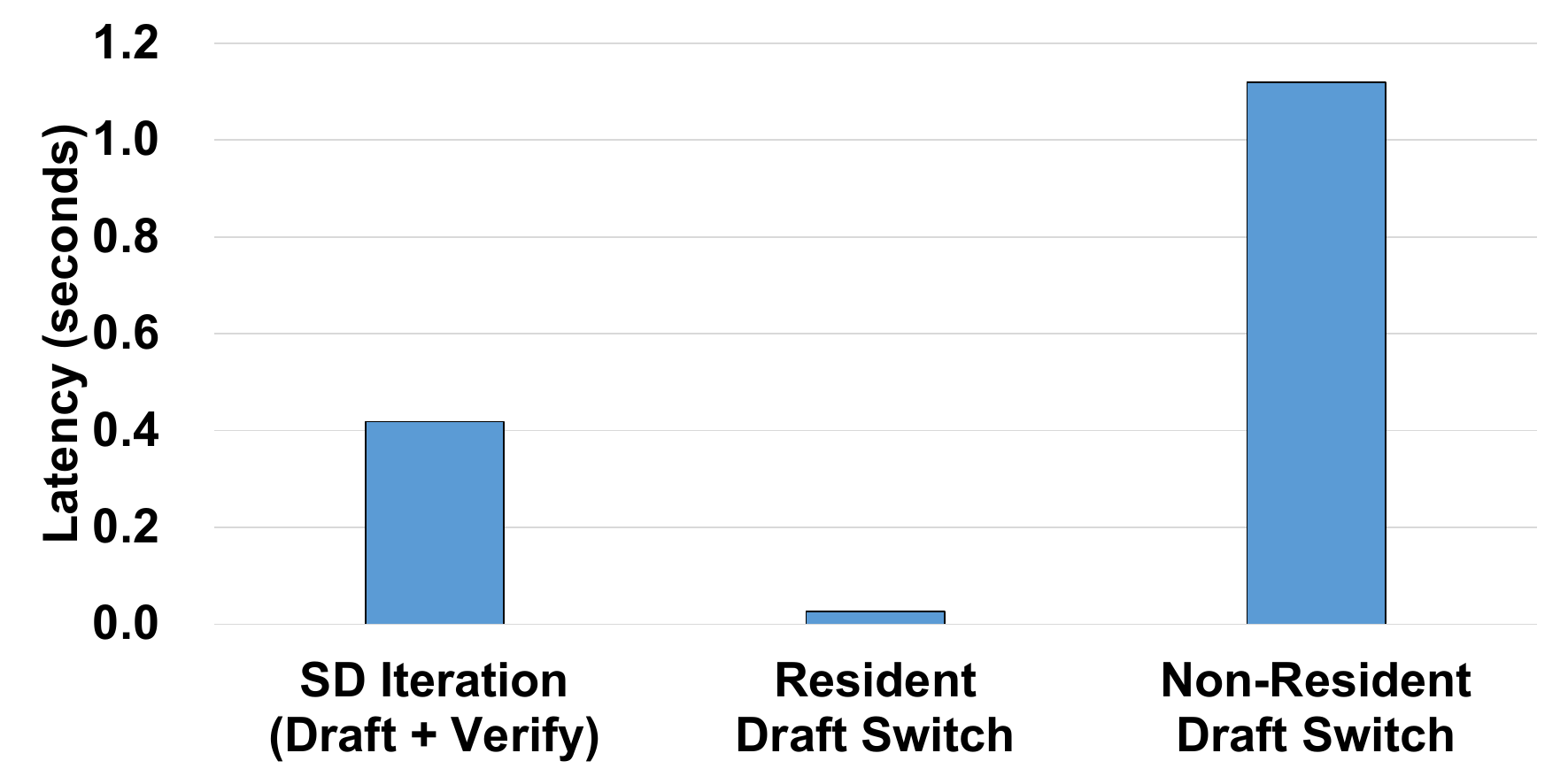}
\caption{Latency comparison between SD iteration and draft switching overheads.}
\label{fig:latency_breakdown}
\end{figure}

We quantify this overhead on a Jetson Orin Nano using a 400M-parameter draft
model. Figure~\ref{fig:latency_breakdown} shows that loading a non-resident
draft takes $2.7\times$ longer than a single SD iteration. As a result, even
infrequent switching can offset the throughput gains from improved acceptance.

\textbf{Observation 2: On memory-constrained edge devices, switching cost
fundamentally limits adaptive draft selection.} Because draft loading latency
significantly exceeds per-iteration decoding latency, switching introduces
substantial overhead. As the number of candidate drafts increases, cache misses
become more frequent, further amplifying this cost.

These results reveal a fundamental trade-off: while adaptive draft selection
can improve acceptance, switching to non-resident drafts incurs high loading
latency that can negate throughput gains. Consequently, naive adaptive
switching strategies are ineffective on memory-constrained edge devices.

\subsection{Limitations of Exploration-Based Adaptive Draft Selection}

\begin{figure}[t]
\centering
\includegraphics[width=\linewidth]{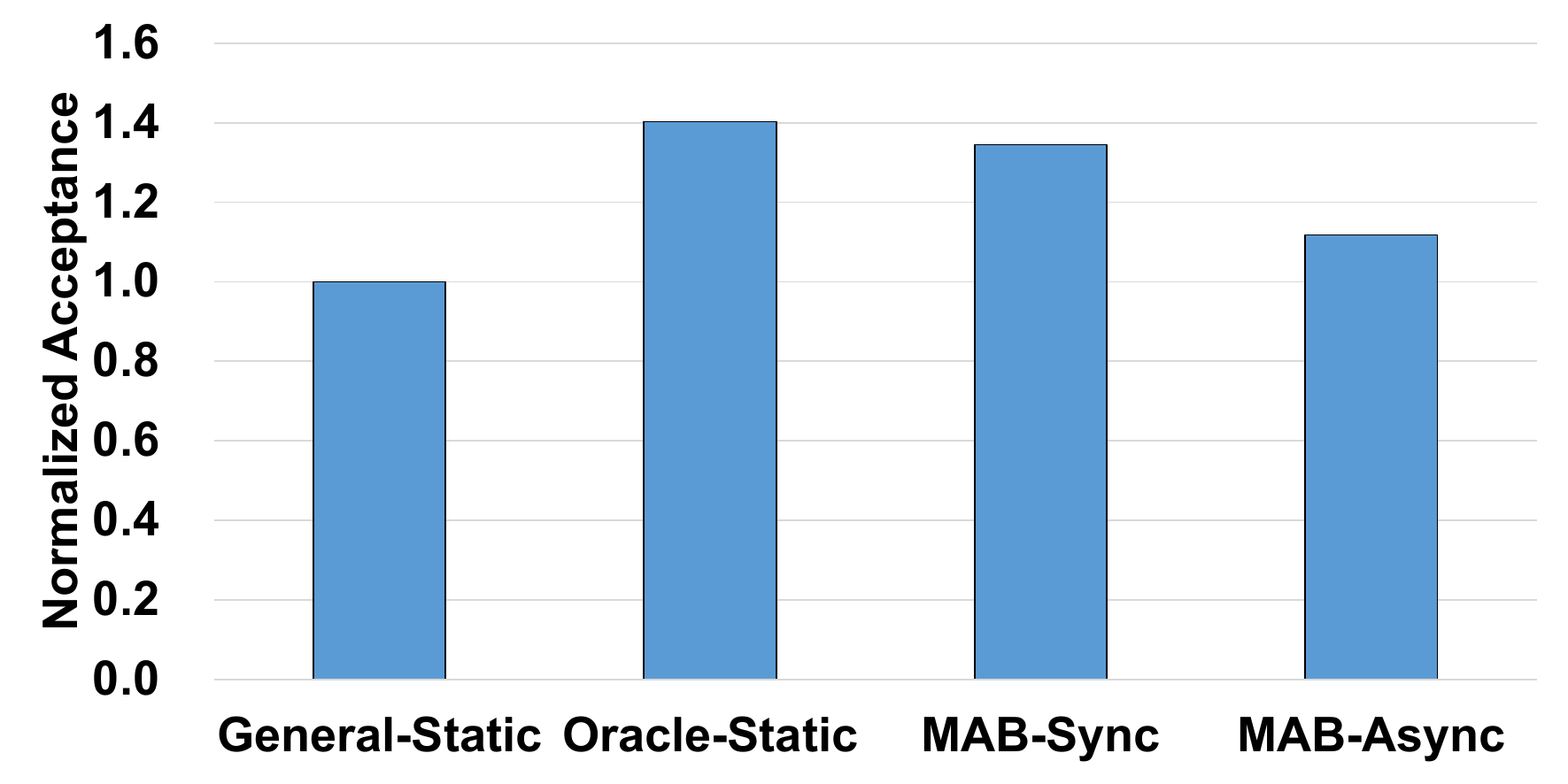}
\caption{Normalized token acceptance of adaptive draft selection methods.}
\label{fig:mab_acceptance}
\end{figure}

\begin{figure}[t]
\centering
\includegraphics[width=\linewidth]{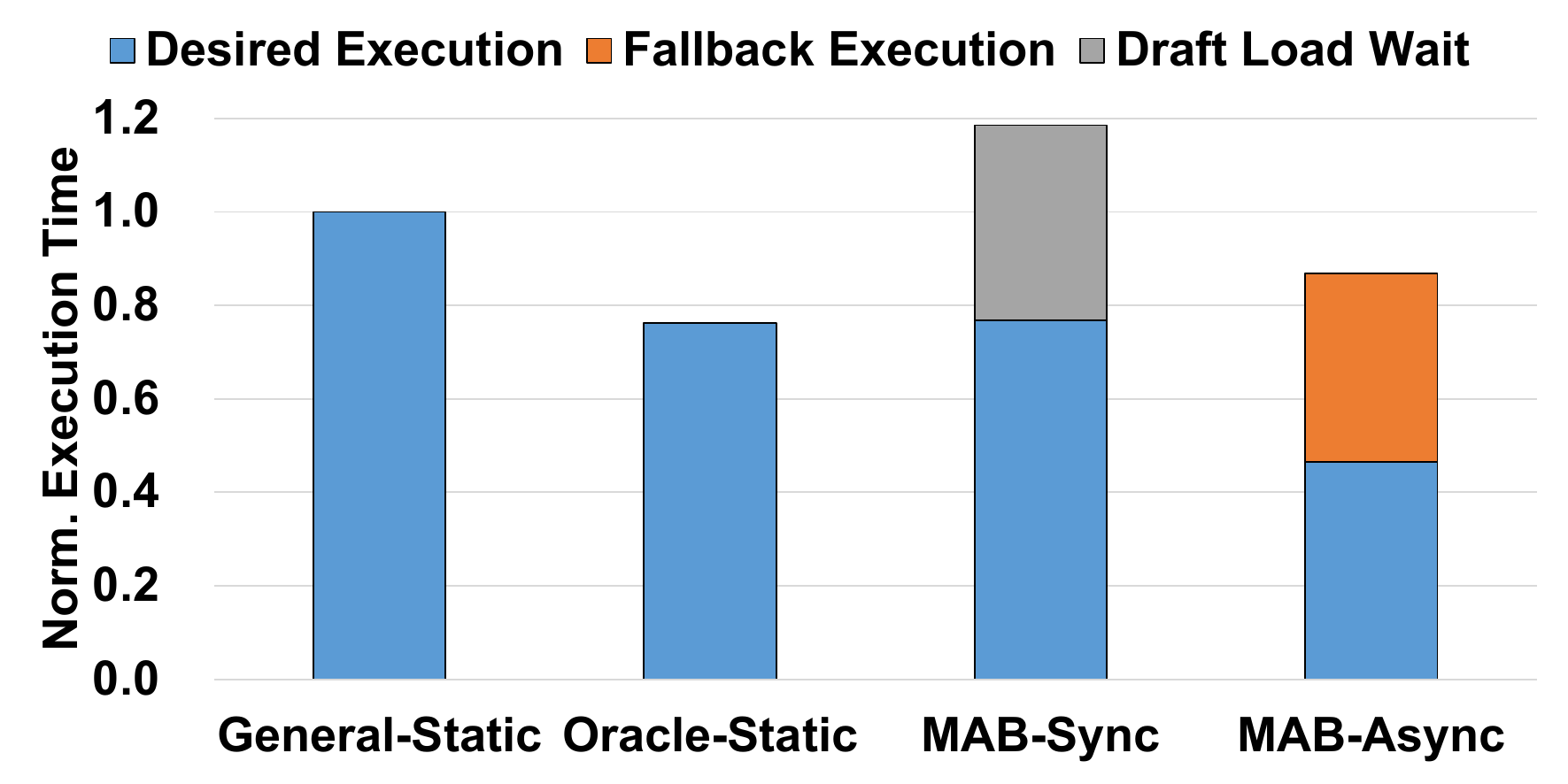}
\caption{Execution time breakdown of exploration-based methods.}
\label{fig:mab_breakdown}
\end{figure}

State-of-the-art adaptive speculative decoding
methods~\cite{hou2025banditspec,kim2025a,liu2026notabandit} rely on multi-armed
bandit (MAB)-based online exploration to identify effective draft models during
generation. These approaches maintain multiple candidate drafts and adaptively
select among them based on observed runtime feedback. In \textit{MAB-Sync}, the
runtime blocks to load and evaluate candidate drafts during exploration,
incurring high switching overhead. In contrast, \textit{MAB-Async} overlaps
model loading with ongoing decoding to reduce blocking, but continues execution
with suboptimal drafts while waiting for preferred drafts to become available.

Figure~\ref{fig:mab_acceptance} shows that \textit{MAB-Sync} improves normalized
acceptance by $34.5\%$ on average over \textit{General-Static}, approaching
\textit{Oracle-Static}. This confirms that exploration-based adaptation can
effectively improve draft quality.

However, these gains do not translate into throughput on memory-constrained
edge devices. To understand why, we analyze execution time breakdown.
Figure~\ref{fig:mab_breakdown} shows that model loading dominates execution
time in \textit{MAB-Sync}. Although \textit{MAB-Async} partially overlaps loading
with execution, it still spends a significant fraction of time—$46.4\%$ on
average—executing suboptimal drafts while waiting for preferred drafts to become
resident.

\begin{figure*}[t]
\centering
\includegraphics[width=0.62\textwidth]{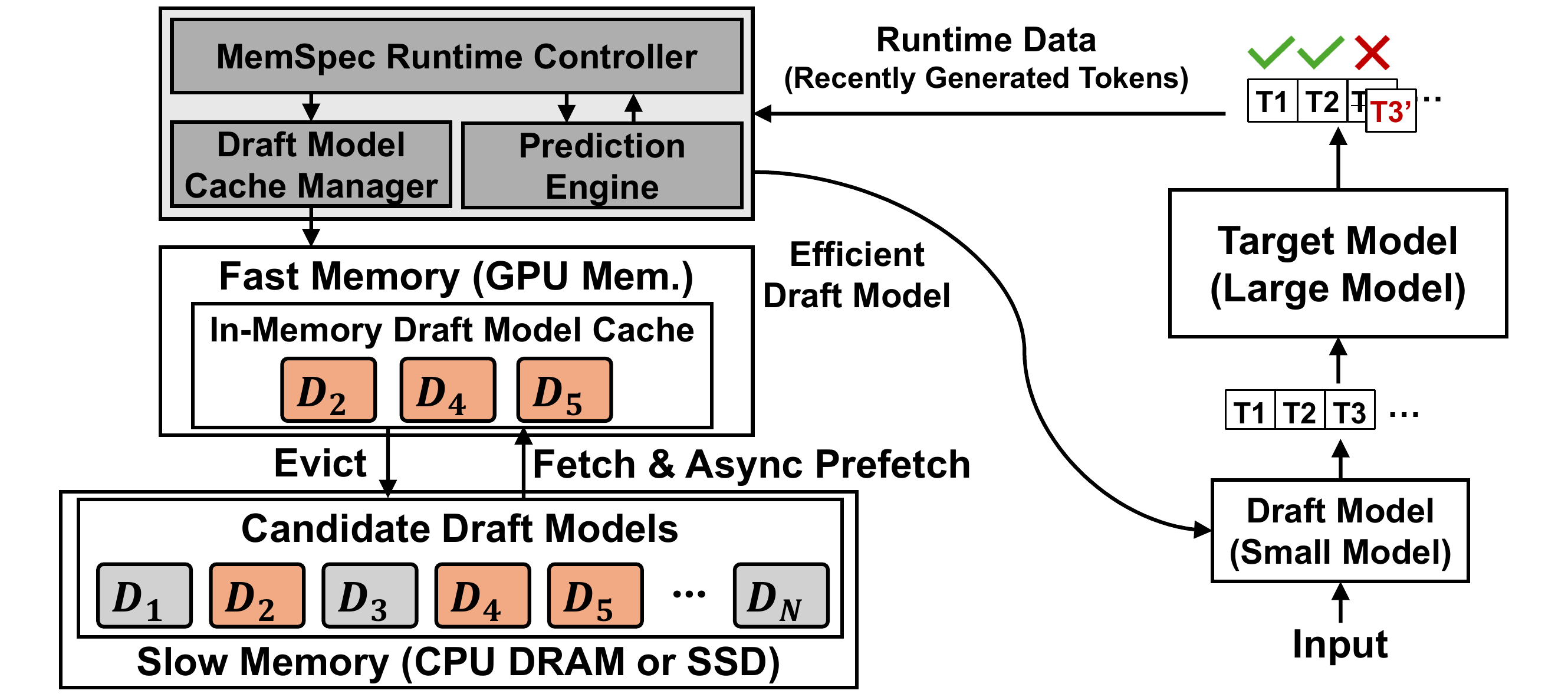}
\caption{Overall architecture of \textsc{MemSpec}.}
\label{fig:memspec-arch}
\end{figure*}

\textbf{Observation 3: Exploration-based adaptation fails to improve throughput under memory constraints.}
While exploration improves draft selection, it incurs frequent model loading,
which is prohibitively expensive on edge devices. As a result, improved
acceptance does not translate into improved throughput.

These results highlight a key limitation of existing adaptive methods: they
optimize \emph{which} draft to use, but ignore \emph{whether} the selected draft
is immediately executable. Under tight memory constraints, this mismatch leads
to excessive fallback execution and diminished performance gains.

Taken together, these observations show that improving acceptance alone is
insufficient. An effective runtime must both identify promising drafts without
repeated exploration and ensure their availability at runtime. This insight
motivates \textsc{MemSpec}, a prediction-guided, memory-aware runtime for
adaptive draft scheduling.

\section{MemSpec Design}
\label{sec:design}

Figure~\ref{fig:memspec-arch} presents the overall architecture of
\textsc{MemSpec}, a prediction-guided, memory-aware runtime for adaptive
draft scheduling in speculative decoding on edge devices. The runtime
integrates three key components: a Prediction Engine that ranks candidate
drafts based on decoding context, a Draft Model Cache Manager that maintains
a small resident working set under memory constraints, and a Runtime
Controller that orchestrates non-blocking adaptive decoding.

The key goal of \textsc{MemSpec} is to realize the benefits of adaptive
draft selection without incurring the high overhead of model switching under
tight memory constraints. Rather than relying on online exploration,
\textsc{MemSpec} predicts a small set of promising draft models for the
current decoding context and maintains them as a resident working set. At
runtime, decoding always proceeds with the best currently resident draft,
while high-priority non-resident drafts are prefetched in the background.

The central challenge is that, on memory-constrained edge devices, draft
selection and execution are no longer tightly coupled: a draft predicted to
be effective may not be immediately executable because it is not resident.
\textsc{MemSpec} addresses this mismatch by formulating adaptive speculative
decoding as a memory-aware runtime scheduling problem, where prediction
determines which drafts should be prepared, while execution is restricted to
drafts that are currently available.

This design leads to two key principles: (1) \emph{non-blocking execution},
which avoids stalling on non-resident drafts by always selecting from the
resident set, and (2) \emph{proactive residency management}, which aligns the
working set of drafts with predicted future demand. Together, these
mechanisms allow \textsc{MemSpec} to achieve efficient adaptive decoding
without repeated model switching or exploration overhead.

\subsection{Design Overview}
\label{subsec:overview}

Consider a speculative decoding run with candidate draft set $\mathcal{D}$ and
resident cache capacity $K$, where at most $K$ draft models can remain
resident in fast memory. Let $\mathcal{G}_i \subseteq \mathcal{D}$ denote the
resident set at scheduling point $i$, and let $d_i \in \mathcal{G}_i$ denote
the active draft used during interval $i$.

\textsc{MemSpec} performs scheduling every $N$ speculative decoding iterations.
This interval-based design serves two purposes: (1) it amortizes the cost of
prediction and runtime control over multiple decoding steps, and (2) it enables
overlap between decoding and asynchronous loading of non-resident drafts,
which is critical for hiding model loading latency.

At initialization, \textsc{MemSpec} selects the first draft using only the
input prompt:
\[
  d_0 = \arg\max_{d \in \mathcal{D}} P(d \mid x_{\text{prompt}}).
\]

During generation, the context at scheduling point $i$ is defined as
\[
  x_i = [\,x_{\text{prompt}}; x^{\text{recent}}_i(T)\,],
\]
where $x^{\text{recent}}_i(T)$ denotes the most recent $T$ generated tokens.
The Prediction Engine assigns each draft a score:
\[
  \boldsymbol{p}_i = \{ P(d \mid x_i) \mid d \in \mathcal{D} \}.
\]

These scores are used in two ways. First, \textsc{MemSpec} derives a ranked
list of candidate drafts:
\[
  R_i = \operatorname{Sort}(\boldsymbol{p}_i),
\]
from which the cache manager derives the target resident set
\[
  \mathcal{W}_i = \operatorname{TopK}(R_i).
\]
Second, it selects the active draft only from the currently resident set:
\[
  d_i^\star = \arg\max_{d \in \mathcal{G}_i} P(d \mid x_i).
\]

The key design abstraction of \textsc{MemSpec} is the separation between the
target working set $\mathcal{W}_i$ and the active draft $d_i^\star$.
Prediction produces a ranked list of candidate drafts, the cache manager
prepares a small working set from that list, and execution always proceeds
with the best currently resident draft.

This design enables non-blocking decoding while gradually steering the cache
toward more effective drafts. Conceptually, \textsc{MemSpec} balances two
competing objectives: (1) following the most promising draft as the generation
context evolves, and (2) minimizing costly model loading under a tight memory
budget. The interval-based working-set design reconciles these objectives by
transforming draft adaptation into a staged, overlapped process, rather than a
sequence of stall-heavy immediate switches.

\subsection{Prediction Engine}
\label{subsec:pred-engine}

The Prediction Engine ranks candidate draft models for the current decoding
context. Unlike exploration-based approaches, \textsc{MemSpec} uses an
offline-trained model to directly predict context-to-draft matching, avoiding
costly online probing of multiple drafts.

We use a fine-tuned BERT encoder as the predictor:
\[
\boldsymbol{p}_i = f_\theta(x_i),
\]
where $\boldsymbol{p}_i = [p_i(d_1), p_i(d_2), \dots, p_i(d_{|\mathcal{D}|})]$
denotes the predicted score for each draft.

The input $x_i$ combines the prompt and recent output tokens. The prompt
captures global task semantics, while recent tokens reflect phase-dependent
generation behavior (e.g., reasoning vs.\ code generation). Using only the
prompt fails to capture such phase transitions, whereas relying only on recent
tokens loses global task intent. Their combination enables effective
context-aware draft ranking.

Importantly, the Prediction Engine is designed to be lightweight.
\textsc{MemSpec} does not require fine-grained utility estimation or online
evaluation of multiple drafts. Instead, it only needs a ranking that is
sufficiently accurate to identify a small set of promising candidates. This is
sufficient because the downstream cache manager and runtime controller operate
on relative priority rather than exact utility values.

\paragraph*{Output and training.}
The predictor outputs a probability for each draft:
\[
p_i(d) = P(d \mid x_i), \quad \forall d \in \mathcal{D}.
\]
It is trained offline using speculative decoding traces, where each context
$x_i$ is labeled with the draft that achieves the highest decoding utility.
This aligns prediction targets with system-level performance.

\paragraph*{Runtime usage.}

\begin{algorithm}[t]
\caption{Context-Aware Draft Ranking}
\label{alg:pred-engine}
\begin{algorithmic}[1]
\State \textbf{Inputs:} prompt $x_{\text{prompt}}$, recent tokens $x^{\text{recent}}_i(T)$, candidate set $\mathcal{D}$
\State \textbf{Output:} ranked list $R_i$

\If{$i = 0$}
    \State $\boldsymbol{p}_i \gets f_\theta(x_{\text{prompt}})$
\Else
    \State $\boldsymbol{p}_i \gets f_\theta([x_{\text{prompt}}; x^{\text{recent}}_i(T)])$
\EndIf

\State sort $d \in \mathcal{D}$ by $p_i(d)$ to obtain $R_i$
\State \Return $R_i$
\end{algorithmic}
\end{algorithm}

The Prediction Engine is invoked once every $N$ iterations. When invoked, it
computes a ranked list of candidate drafts as summarized in
Algorithm~\ref{alg:pred-engine}. Given the current context, it computes the
per-draft score vector (lines 4--6), sorts drafts by score to obtain the ranked
list $R_i$ (line 7), and returns this list for downstream cache management (line
8). The target working set $\mathcal{W}_i$ is then derived by the cache manager
from the top-$K$ drafts in $R_i$. Because prediction is performed only at
interval boundaries, its overhead is amortized over multiple decoding steps.

\subsection{Draft Model Cache Manager}
\label{subsec:cache-manager}

The Draft Model Cache Manager maintains draft residency under a tight memory
budget. Because only $K$ drafts can reside in fast memory, \textsc{MemSpec}
explicitly manages a small working set instead of keeping all candidate drafts
loaded.

We consider two memory tiers: (1) a \emph{resident draft cache} containing
immediately executable drafts, and (2) \emph{backing storage} containing drafts
that require asynchronous loading.

\begin{algorithm}[t]
\caption{Draft Model Cache Update}
\label{alg:cache-manager}
\begin{algorithmic}[1]
\State \textbf{Inputs:} $R_i$, $\mathcal{G}_i$, capacity $K$, active draft $d_i$

\State $\mathcal{W}_i \gets$ top-$K$ drafts in $R_i$

\For{each $d \in \mathcal{W}_i$}
    \If{$d \notin \mathcal{G}_i$ and not being prefetched}
        \While{$|\mathcal{G}_i| = K$}
            \State select $d_{\text{evict}} \in \mathcal{G}_i \setminus (\mathcal{W}_i \cup \{d_i\})$ with lowest score
            \If{no such draft exists}
                \State \textbf{break}
            \EndIf
            \State evict $d_{\text{evict}}$
        \EndWhile
        \If{$|\mathcal{G}_i| < K$}
            \State asynchronously prefetch $d$
        \EndIf
    \EndIf
\EndFor
\end{algorithmic}
\end{algorithm}

At each scheduling point, the cache manager derives the target set
$\mathcal{W}_i$ from the ranked list $R_i$ by selecting the top-$K$ drafts
(Algorithm~\ref{alg:cache-manager}, line 2). It then incrementally updates the
resident set $\mathcal{G}_i$ without blocking execution.

\paragraph*{Policy.}
For each $d \in \mathcal{W}_i \setminus \mathcal{G}_i$, the system initiates
asynchronous prefetch (lines 3--11). If space is required, it evicts drafts in
$\mathcal{G}_i \setminus \mathcal{W}_i$ in ascending order of predicted utility,
while protecting the active draft $d_i$ whenever possible (lines 5--9). This
simple top-$K$ policy is effective in practice, as prediction already captures
most of the utility variation across drafts.

This design decouples long-horizon planning from immediate execution. Rather
than switching to the globally best draft immediately, \textsc{MemSpec}
incrementally reshapes the resident set so that promising drafts become
available at future scheduling points. In this sense, cache management is not
merely a storage optimization, but a key component of adaptive draft scheduling.

\subsection{Runtime Controller}
\label{subsec:runtime-controller}

The Runtime Controller orchestrates decoding, prediction, cache updates, and
draft switching. Its core policy is to always execute the best currently
resident draft and never stall for non-resident models.

\begin{algorithm}[t]
\caption{\textsc{MemSpec} Runtime Loop}
\label{alg:runtime}
\begin{algorithmic}[1]
\State load target model
\State $R_0 \gets \textsc{PredictDraftRanking}(x_{\text{prompt}})$
\State load initial draft $d_0$
\State $\mathcal{G}_0 \gets \{d_0\}$
\State \textsc{UpdateCache}($R_0, \mathcal{G}_0, K, d_0$)
\State $i \gets 0$

\While{not end-of-sequence}
    \For{$j = 1$ to $N$}
        \State run draft $d_i$
        \State verify with target model
        \If{end-of-sequence}
            \State \textbf{break}
        \EndIf
    \EndFor

    \If{not end-of-sequence}
        \State collect $x^{\text{recent}}_i(T)$
        \State $x_{i+1} \gets [x_{\text{prompt}}; x^{\text{recent}}_i(T)]$
        \State $R_{i+1} \gets \textsc{PredictDraftRanking}(x_{i+1})$
        \State \textsc{UpdateCache}($R_{i+1}, \mathcal{G}_i, K, d_i$)
        \State $d_{i+1} \gets \arg\max_{d \in \mathcal{G}_i} P(d \mid x_{i+1})$
        \State $i \gets i + 1$
    \EndIf
\EndWhile
\end{algorithmic}
\end{algorithm}

As summarized in Algorithm~\ref{alg:runtime}, \textsc{MemSpec} first predicts an
initial ranked list from the prompt and loads the initial draft (lines 2--4).
It then initializes the resident set and triggers cache preparation (line 5).

During execution, the controller repeatedly performs three steps. First, it runs
speculative decoding for the current interval using the active draft $d_i$
(lines 8--12). Second, it constructs the next context from the prompt and the
most recent generated tokens (lines 14--15). Third, it invokes prediction and
cache update to prepare future drafts, and selects the best currently resident
draft for the next interval (lines 16--18).

If the top-ranked draft is not yet resident, execution continues with the best
available draft. Once loading completes, the new draft becomes eligible at the
next scheduling point. This ensures that incomplete prefetch does not introduce
blocking and converts draft adaptation into an overlapped process.

This policy captures the key principle of \textsc{MemSpec}: the runtime should
never sacrifice immediate progress to follow an unavailable draft. Instead, it
continues decoding with the best runnable draft while opportunistically
preparing better candidates in the background. This best-effort switching
strategy enables adaptive decoding without incurring the severe overhead of
naive dynamic switching.

\subsection{System Implementation}
\label{subsec:implementation}

\textsc{MemSpec} is implemented in PyTorch and built on top of a
high-performance speculative decoding engine presented in~\cite{NEURIPS2024_1d91d568}.
All models share a common tokenizer, and draft execution follows the standard
draft-and-verify process within a conventional speculative decoding pipeline.

Each draft model maintains its residency state and prefetch status.
Prefetching is performed asynchronously and overlapped with decoding,
allowing model loading to proceed without blocking execution. The active
draft is protected from eviction whenever possible to ensure uninterrupted
decoding.

The Prediction Engine is invoked once every $N$ iterations, amortizing its
runtime overhead. In practice, prediction accounts for only 3.9\% of total
execution time on average.

Overall, \textsc{MemSpec} enables adaptive draft selection as a lightweight
and non-blocking runtime mechanism, making it well-suited for
memory-constrained edge deployments.

\section{Evaluation}
\label{sec:eval}

We evaluate \textsc{MemSpec} along four key aspects:

\begin{itemize}[noitemsep,nolistsep]
\item \textbf{Generation throughput:} How much steady-state generation
throughput improvement can \textsc{MemSpec} achieve over static and adaptive
draft-selection strategies?

\item \textbf{Performance breakdown:} Why does \textsc{MemSpec} outperform
exploration-based adaptive methods on memory-constrained edge devices?

\item \textbf{Impact of design components:} How much do the key design
components of \textsc{MemSpec} contribute to performance?

\item \textbf{Sensitivity analysis:} How sensitive is \textsc{MemSpec} to
runtime parameters such as scheduling interval, output length, and resident
cache capacity?

\end{itemize}

\subsection{Experimental Setup}
\label{subsec:eval-setup}

\noindent\textbf{Platform.}
We evaluate \textsc{MemSpec} on a Jetson Orin Nano platform. Table~\ref{tab:hw}
summarizes the hardware configuration.

\begin{table}[t]
\centering
\small
\caption{Hardware configuration.}
\label{tab:hw}

\begin{tabular}{@{}lp{0.7\linewidth}@{}}
\toprule
\textbf{Component} & \textbf{Specification} \\
\midrule
Platform & NVIDIA Jetson Orin Nano \\
GPU & Ampere, 1024 CUDA cores, 32 Tensor cores \\
CPU & 6-core Arm Cortex-A78AE \\
Memory & 8GB LPDDR5 (102 GB/s) \\
Storage & 1TB Samsung 990 PRO M.2 NVMe SSD \\
\bottomrule
\end{tabular}
\end{table}

\noindent\textbf{Software.}
We use PyTorch 2.9.1 with CUDA 12.6 on NVIDIA JetPack 6.2.1.
All experiments are conducted under the same software environment.

\noindent\textbf{Models.}
We evaluate two target models: GPTQ INT4 \textsc{LLaMA-2} 7B~\cite{touvron2023llama2openfoundation}
and GPTQ INT4 \textsc{Qwen2.5} 7B~\cite{qwen2025qwen25technicalreport}.
For each model family, we construct five draft models: one general-purpose draft
and four domain-specialized drafts (code, math, law, and medical).
Each draft model has 400M parameters for LLaMA-2 and 0.5B parameters for Qwen2.5.

Under the 8GB memory constraint of the Jetson Orin Nano, at most two draft
models can remain resident simultaneously. Therefore, we use $K=2$ in the
main evaluation unless otherwise stated.

\noindent\textbf{Draft models.} Domain-specialized drafts are obtained via
distillation following prior work~\cite{yi-etal-2024-towards}. Each draft is
fine-tuned on domain-specific datasets:
GSM8K~\cite{cobbe2021trainingverifierssolvemath} and
MATH~\cite{hendrycks2021measuring} (math reasoning),
HumanEval~\cite{chen2021evaluating} and
MBPP~\cite{austin2021programsynthesislargelanguage} (code),
Lex\-GLUE~\cite{chalkidis-etal-2022-lexglue} (law), and MedQA~\cite{app11146421}
and MedMCQA~\cite{pmlr-v174-pal22a} (medical).

\noindent\textbf{Prediction model.}
The Prediction Engine uses a BERT-based model with a task-specific ranking
head. The encoder is kept fixed, and only the ranking head is trained.
Training data is collected by running speculative decoding on the same
datasets used for draft fine-tuning and recording execution traces. The
predictor is trained to select the most effective draft for each decoding
context based on observed runtime utility. Training data is disjoint from
the evaluation datasets to avoid data leakage and assess generalization.

\noindent\textbf{Workloads.}
We evaluate on five datasets covering diverse domains: Alpaca~\cite{alpaca}
(instruction following), LiveCodeBench\-~\cite{jain2025livecodebench} (code),
Omni-MATH~\cite{gao2024omnimathuniversalolympiadlevel} (math),
MMLU-Law, and MMLU-Medical~\cite{hendrycks2021measuring} (MMLU professional law
and medicine). We randomly sample 100 prompts from each dataset.

\noindent\textbf{Runtime configuration.}
Unless otherwise stated, the scheduling interval is set to $N=4$ and the output
length is fixed to 128 generated tokens. We use greedy decoding with batch size
1.

\noindent\textbf{Throughput measurement.} 
Throughput is measured as generated tokens per second during steady-state
iterative decoding, including both draft execution and target verification. We
exclude prompt encoding time and focus on steady-state generation so that the
reported performance reflects the efficiency of runtime draft scheduling
itself. All results are averaged over three runs. For each prompt, we reset the
runtime state, including draft cache contents and prediction context, to avoid
cross-sample interference. When summarizing performance across workloads, we
report the geometric mean (GMEAN).

\noindent\textbf{Baselines.}
We compare the following draft-selection strategies:

\begin{itemize}[noitemsep,nolistsep]
    \item \textbf{General-Static}: uses a single general-purpose (i.e., not fine-tuned) draft model.
    \item \textbf{Oracle-Static}: selects the best draft per input via offline evaluation but keeps it fixed during generation.
    \item \textbf{MAB-Async}: a state-of-the-art exploration-based adaptive method that dynamically selects drafts without blocking execution.
    \item \textbf{Oracle-Dynamic}: an empirical upper bound that selects the best
      draft at each scheduling point using the same scheduling interval as
      \textsc{MemSpec}, with oracle knowledge of future draft utility under the
      same memory constraint.
    \item \textbf{MemSpec}: our prediction-guided, memory-aware runtime.
\end{itemize}

For \textit{MAB-Async}, we implement an asynchronous bandit-based selection
strategy that overlaps draft loading with ongoing decoding. This design gives
the adaptive baseline the benefit of non-blocking loading and therefore
represents a stronger comparison than a synchronous exploration strategy. We use
a UCB-based policy similar to prior bandit-based adaptive speculative decoding
approaches such as BanditSpec~\cite{hou2025banditspec}, and initially explore
each draft model once before adaptive selection begins. During decoding,
MAB-Async adaptively updates draft selection decisions based on observed runtime
behavior. When the selected draft is non-resident, loading proceeds
asynchronously while decoding continues with the currently resident draft. We
evaluate exploration coefficients \((c \in \{0.1, 0.5, 1.0, 2.0, 3.0\})\) and
use the best-performing configuration (\(c=2.0\)) in all experiments.

These baselines cover static, adaptive, and oracle configurations. Comparing
\textit{Oracle-Static} and \textit{Oracle-Dynamic} isolates the benefit of
dynamic adaptation beyond optimal static selection, while comparing
\textit{MAB-Async} and \textsc{MemSpec} highlights the benefit of
prediction-guided, memory-aware scheduling over exploration-based adaptation.

\subsection{Overall Throughput}
\label{subsec:overall-throughput}

\begin{figure*}[t]
\centering
\includegraphics[width=1.0\linewidth]{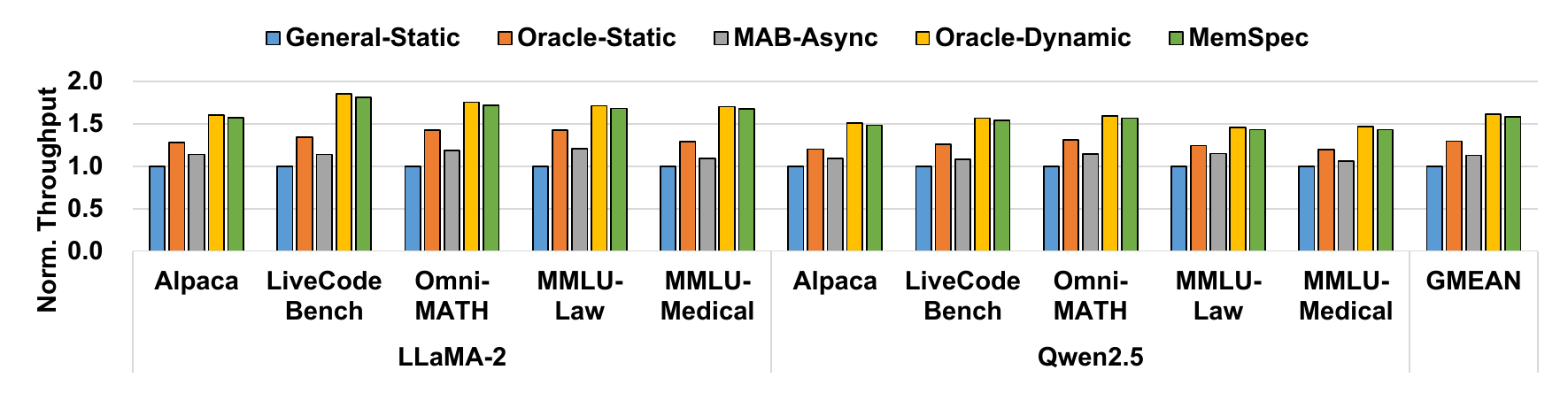}
\caption{Normalized steady-state generation throughput across workloads.}
\label{fig:throughput_main}
\end{figure*}

Figure~\ref{fig:throughput_main} shows normalized steady-state generation
throughput (normalized to \textit{General-Static}) across all workloads for
both \textsc{LLaMA-2} and \textsc{Qwen2.5}. \textsc{MemSpec} consistently
outperforms both static baselines and the adaptive \textit{MAB-Async} method
across all datasets.

\textit{Oracle-Static} improves throughput by $22.5\%$ on average over
\textit{General-Static}, demonstrating that selecting an appropriate draft
model is critical for performance. However, \textit{Oracle-Dynamic} achieves a
further $24.9\%$ improvement over \textit{Oracle-Static}, revealing substantial
additional headroom from dynamic adaptation

\textsc{MemSpec} closely approaches \textit{Oracle-Dynamic}, achieving
$95$--$97\%$ of its throughput while remaining practical for deployment.
Compared with \textit{MAB-Async}, \textsc{MemSpec} improves throughput by
approximately $40.7\%$ on average across workloads.

These results reveal two important insights. First, there exists substantial
dynamic headroom beyond static selection. While \textit{Oracle-Static}
already captures the best per-prompt draft, \textit{Oracle-Dynamic} further
improves performance, indicating that the most effective draft can change
within a single generation. Second, realizing this headroom on edge devices
requires more than better draft identification. Although \textit{MAB-Async}
improves selection quality, it fails to convert much of this benefit into
throughput because effective drafts are often not resident when needed.

\textsc{MemSpec} bridges this gap by aligning draft selection with runtime
availability. By proactively maintaining a small set of high-utility drafts,
it avoids excessive switching while preserving most of the gains of dynamic
adaptation. The remaining gap to \textit{Oracle-Dynamic} is relatively small,
suggesting that prediction-guided scheduling is sufficient to approximate
near-ideal behavior without exhaustive oracle knowledge.

Overall, these results demonstrate that the key bottleneck on edge devices is
not simply identifying better drafts, but ensuring that they are available at
execution time. By jointly optimizing draft selection and draft residency,
\textsc{MemSpec} captures most of the dynamic headroom of
\textit{Oracle-Dynamic} without requiring impractical offline enumeration.

\subsection{Execution Breakdown}
\label{subsec:exec-breakdown}

To understand the performance gap, we decompose execution time into three
components: \emph{desired execution}, \emph{fallback execution}, and
\emph{prediction overhead}. Desired execution corresponds to decoding with the
preferred draft, while fallback execution corresponds to decoding with a
non-optimal resident draft when the desired draft is not yet available.

\begin{figure}[t]
\centering
\includegraphics[width=1.0\linewidth]{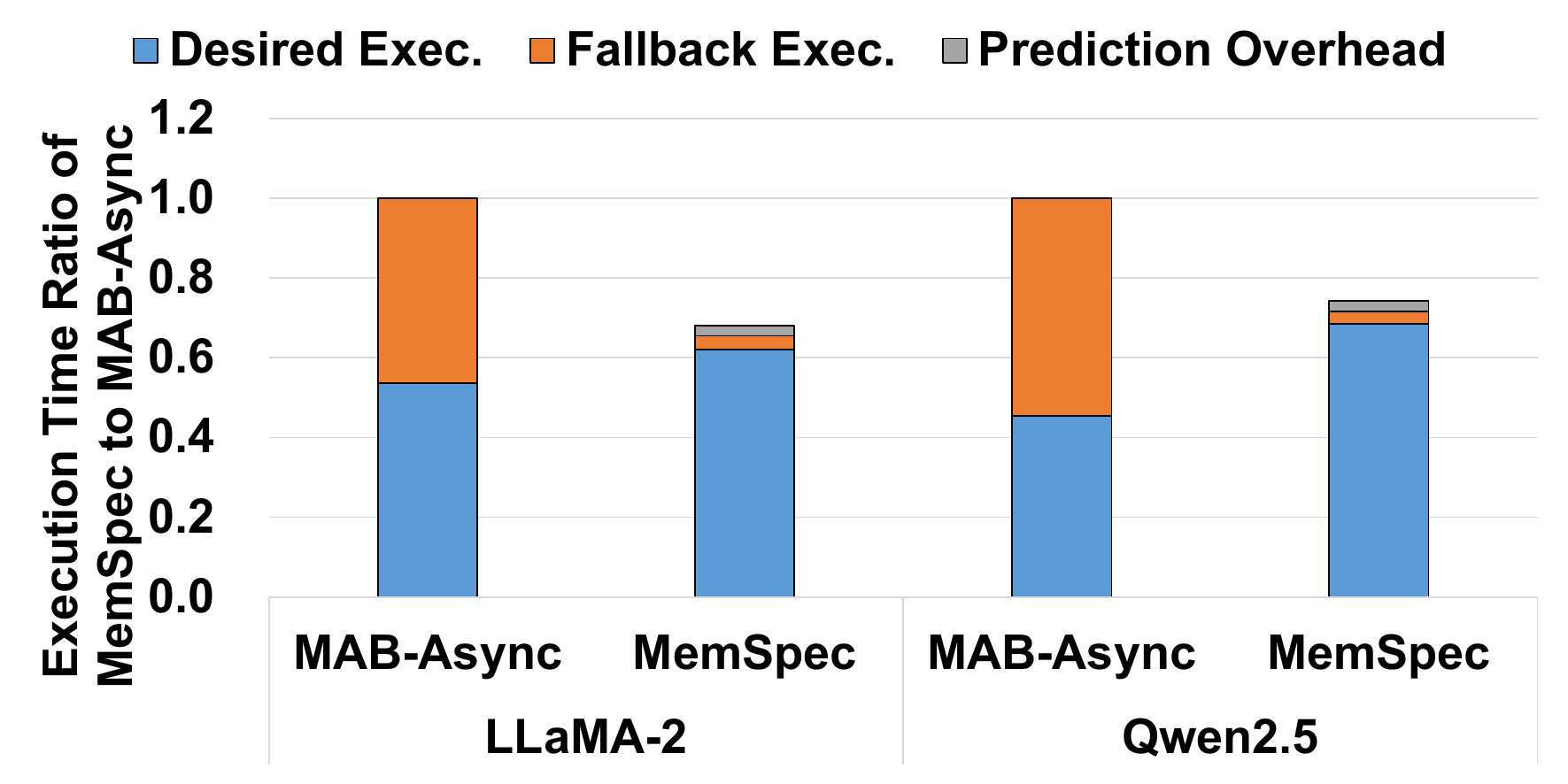}
\caption{Execution time breakdown of \textit{MAB-Async} and \textsc{MemSpec}.}
\label{fig:breakdown_main}
\end{figure}

Figure~\ref{fig:breakdown_main} shows that \textit{MAB-Async} spends a large
fraction of execution time in fallback execution, accounting for $49.4\%$ of
total runtime on average across model families. This reflects a key limitation
of exploration-based adaptation: even when a better draft is identified, it is
often unavailable at execution time, forcing the runtime to continue decoding
with a suboptimal resident draft.

In contrast, \textsc{MemSpec} reduces fallback execution to below $5.5\%$ while
substantially increasing the fraction of desired execution. Prediction overhead
remains negligible, accounting for less than $3.9\%$ of total runtime.

A closer inspection reveals that this reduction is primarily due to improved
temporal alignment between selection and residency. In \textit{MAB-Async},
draft decisions are driven by online exploration and therefore react to past
observations, but do not explicitly prepare high-utility drafts before they
are needed. As a result, the runtime frequently spends time executing
available but inferior drafts while preferred ones are still being loaded.

By contrast, \textsc{MemSpec} uses prediction to anticipate near-future draft
utility from both prompt semantics and recent generation context. This allows
the runtime to initiate prefetch ahead of time and convert reactive switching
into proactive preparation. The resulting increase in desired execution shows
that the performance advantage of \textsc{MemSpec} comes not only from
selecting better drafts, but from ensuring that those drafts are actually
runnable at the scheduling point.

These results highlight a fundamental distinction between \emph{selection}
quality and \emph{execution} quality. While \textit{MAB-Async} can identify
effective drafts, it does not ensure that they are available at execution
time. \textsc{MemSpec} improves throughput by aligning draft selection with
runtime availability through proactive residency management.

\subsection{Impact of Design Components}
\label{subsec:ablation}

We analyze the contribution of \textsc{MemSpec}'s key design components by
comparing the following configurations:

\begin{itemize}[noitemsep,nolistsep]
\item \textbf{General-Static}: a non-adaptive baseline with no prediction or
runtime draft management.
\item \textbf{Prediction-Only}: uses prediction-guided draft selection without
proactive prefetching.
\item \textbf{MemSpec}: the full design with prediction-guided scheduling and
memory-aware prefetching.
\end{itemize}

\begin{figure}[t]
\centering
\includegraphics[width=1.0\linewidth]{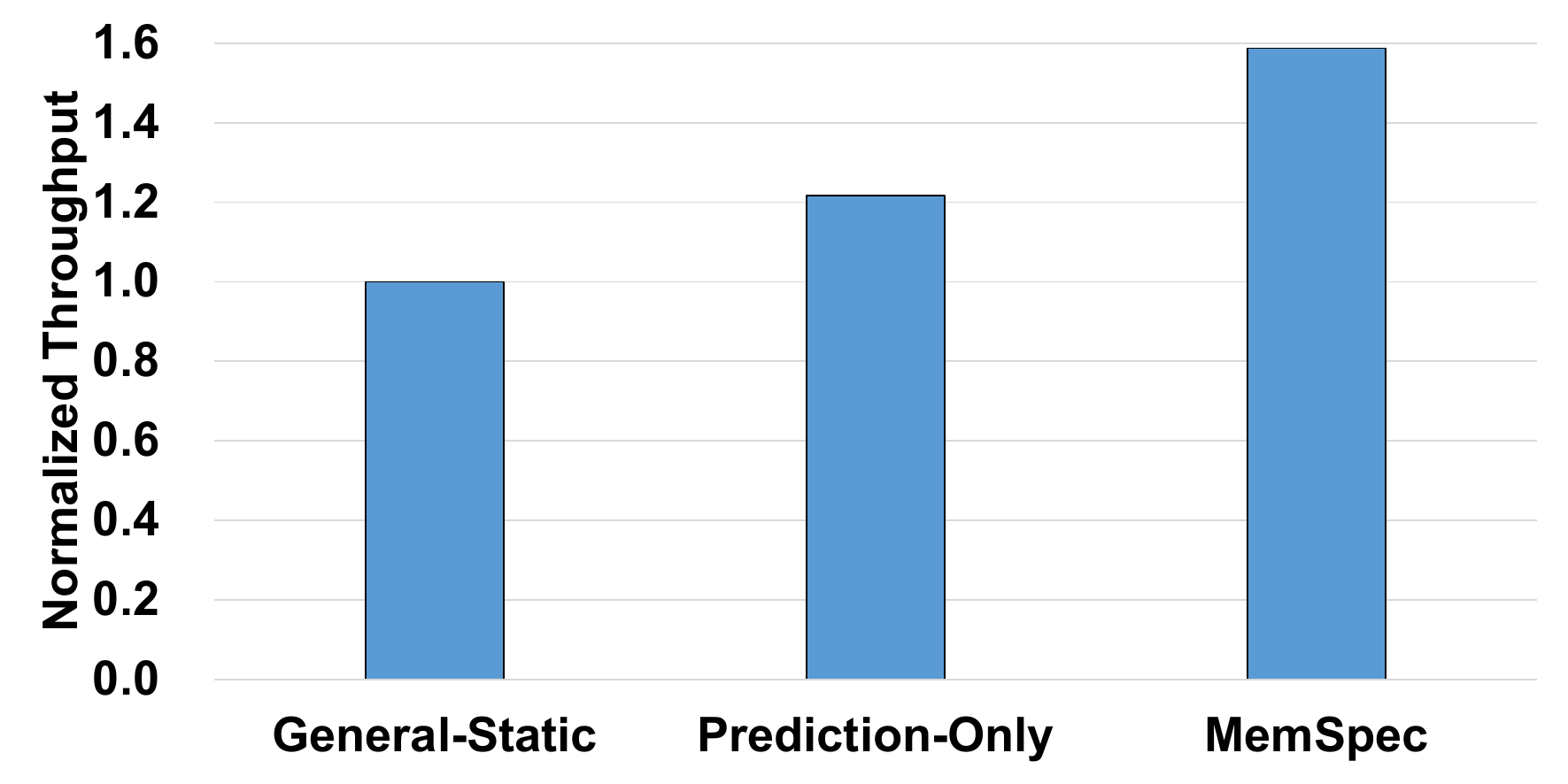}
\caption{Throughput comparison for component analysis.}
\label{fig:ablation_main}
\end{figure}

\begin{figure}[t]
\centering
\includegraphics[width=1.0\linewidth]{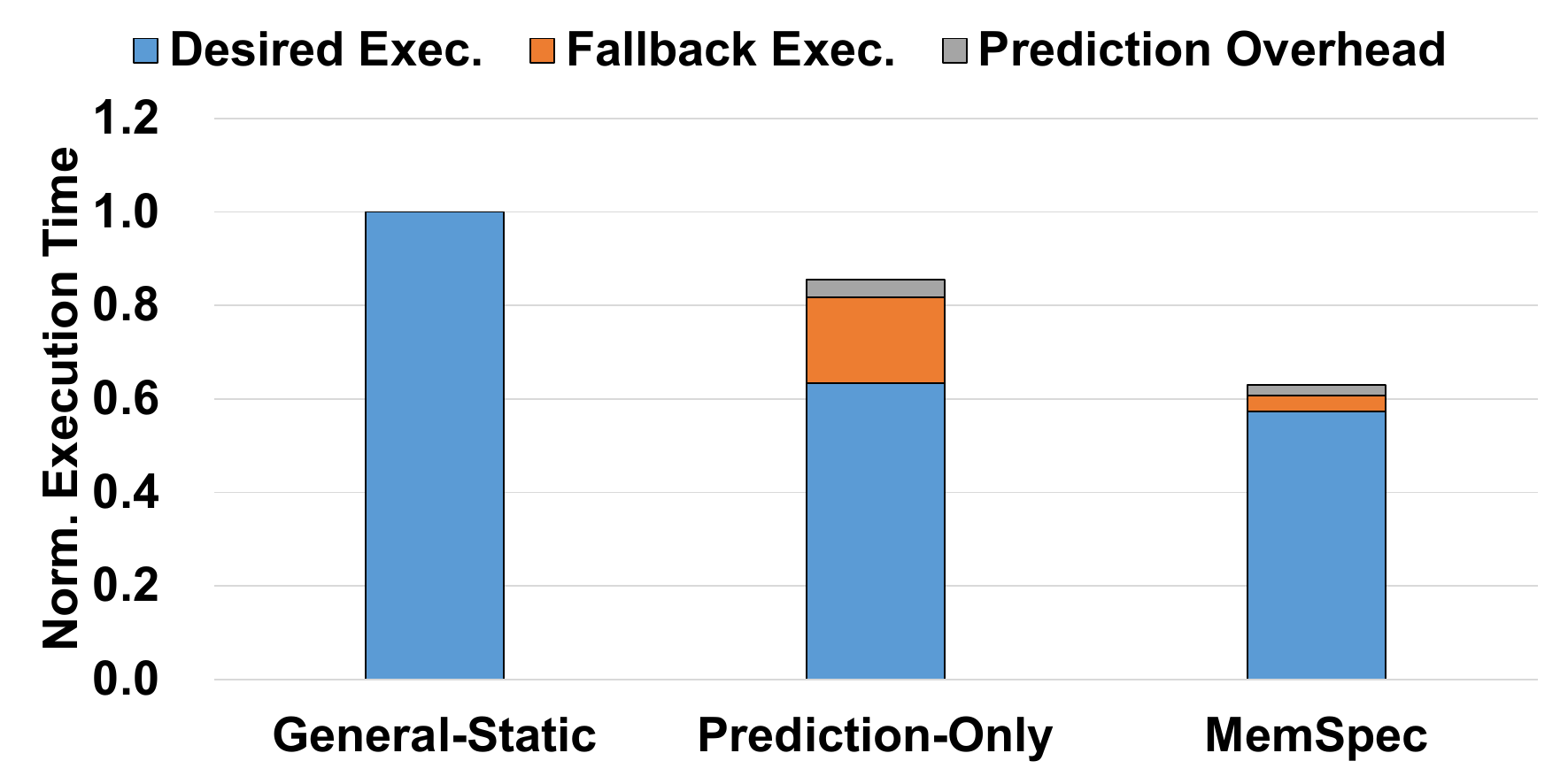}
\caption{Execution breakdown for component analysis.}
\label{fig:ablation_breakdown}
\end{figure}

Figure~\ref{fig:ablation_main} shows that \textit{Prediction-Only} improves
throughput by $21.6\%$ over \textit{General-Static},
confirming that context-aware draft selection is beneficial. However, it still
falls short of \textsc{MemSpec} by $30.5\%$.

Figure~\ref{fig:ablation_breakdown} explains this gap. \textit{Prediction-Only}
spends $21.3\%$ of runtime in fallback execution, while \textsc{MemSpec}
reduces this to below $5.3\%$. This shows that prediction alone is insufficient:
performance gains are realized only when predicted drafts are made available at
execution time.

These results highlight an important system-level insight: prediction accuracy
alone is not the limiting factor in adaptive decoding under memory constraints.
Even when the runtime can identify effective drafts, those drafts may still
fail to improve throughput if they are unavailable when needed. Without
proactive cache management, prediction quality does not directly translate into
execution quality.

\textsc{MemSpec} addresses this limitation by coupling prediction with
residency management, ensuring that high-utility drafts are not only selected
but also prepared in advance. This tight integration between prediction and
scheduling is essential for realizing the full benefit of adaptive decoding.

Overall, \textsc{MemSpec} derives its performance from combining prediction
with memory-aware scheduling, rather than prediction alone. More broadly,
these results show that adaptive speculative decoding on edge devices is
fundamentally a joint optimization problem over draft selection and memory
management.

\subsection{Sensitivity Analysis}
\label{subsec:sensitivity}

We analyze the sensitivity of \textsc{MemSpec} to several key runtime
parameters.

\paragraph*{Scheduling interval.}

\begin{figure}[t]
\centering
\includegraphics[width=1.0\linewidth]{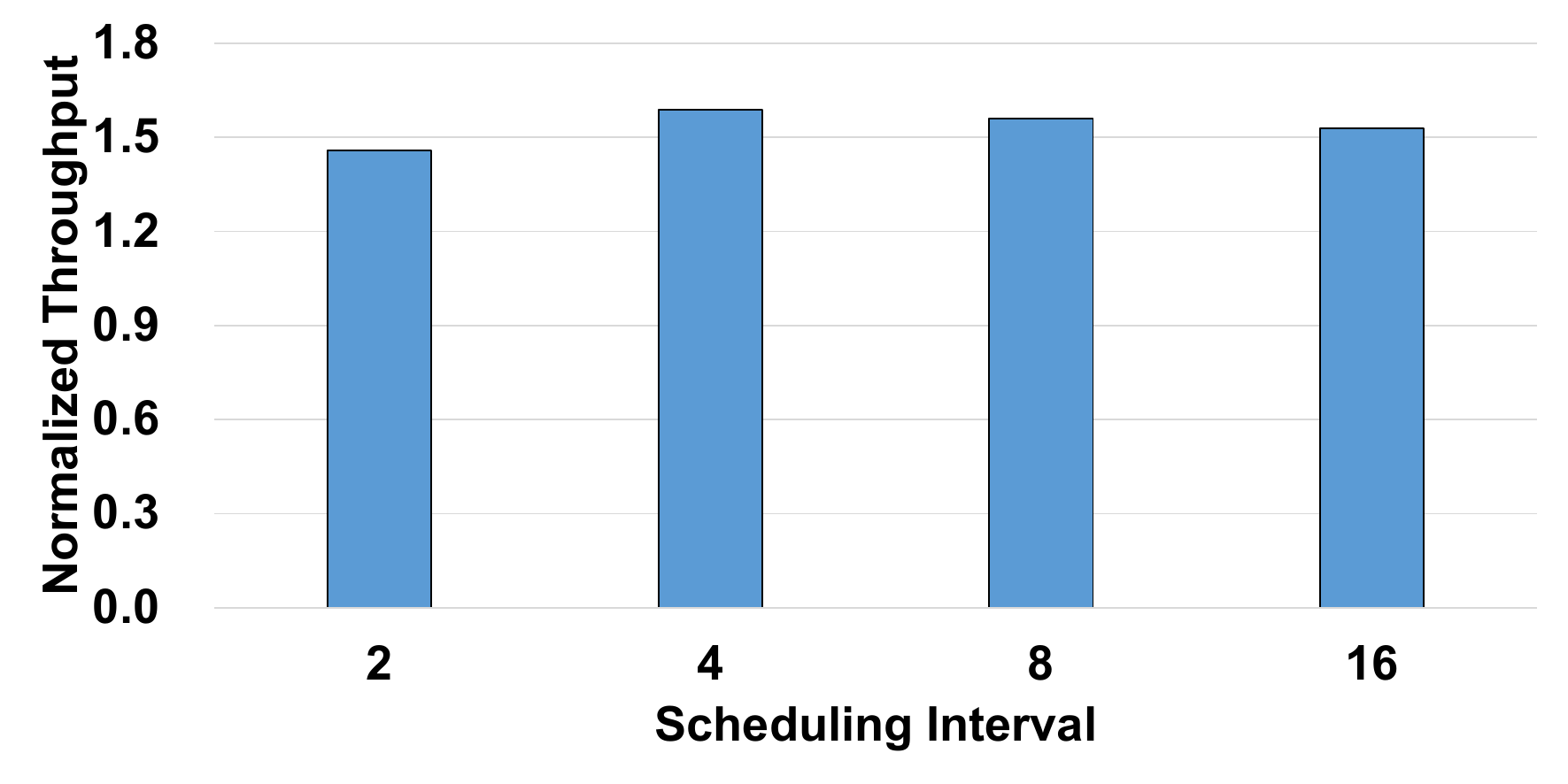}
\caption{Sensitivity to scheduling interval.}
\label{fig:sensitivity_interval}
\end{figure}

Figure~\ref{fig:sensitivity_interval} shows the effect of varying the
scheduling interval. Throughput improves from $N=2$ to $N=4$, remains nearly
unchanged at $N=8$, and decreases at $N=16$. This trend reflects the trade-off
between adaptation frequency and prefetch opportunity. When the interval is too
small, frequent rescheduling leaves less time to overlap model loading with
decoding. When the interval is too large, beneficial draft transitions are
delayed.

The relatively flat performance between $N=4$ and $N=8$ suggests that
\textsc{MemSpec} does not require overly fine-grained parameter tuning to
perform well. This is desirable for practical deployment, where the best
interval may vary slightly across platforms and workloads. Overall, these
results indicate that a moderate interval provides the best balance, and that
\textsc{MemSpec} remains robust over a reasonably wide operating range around
the default setting.

\paragraph*{Output length.}

\begin{figure}[t]
\centering
\includegraphics[width=1.0\linewidth]{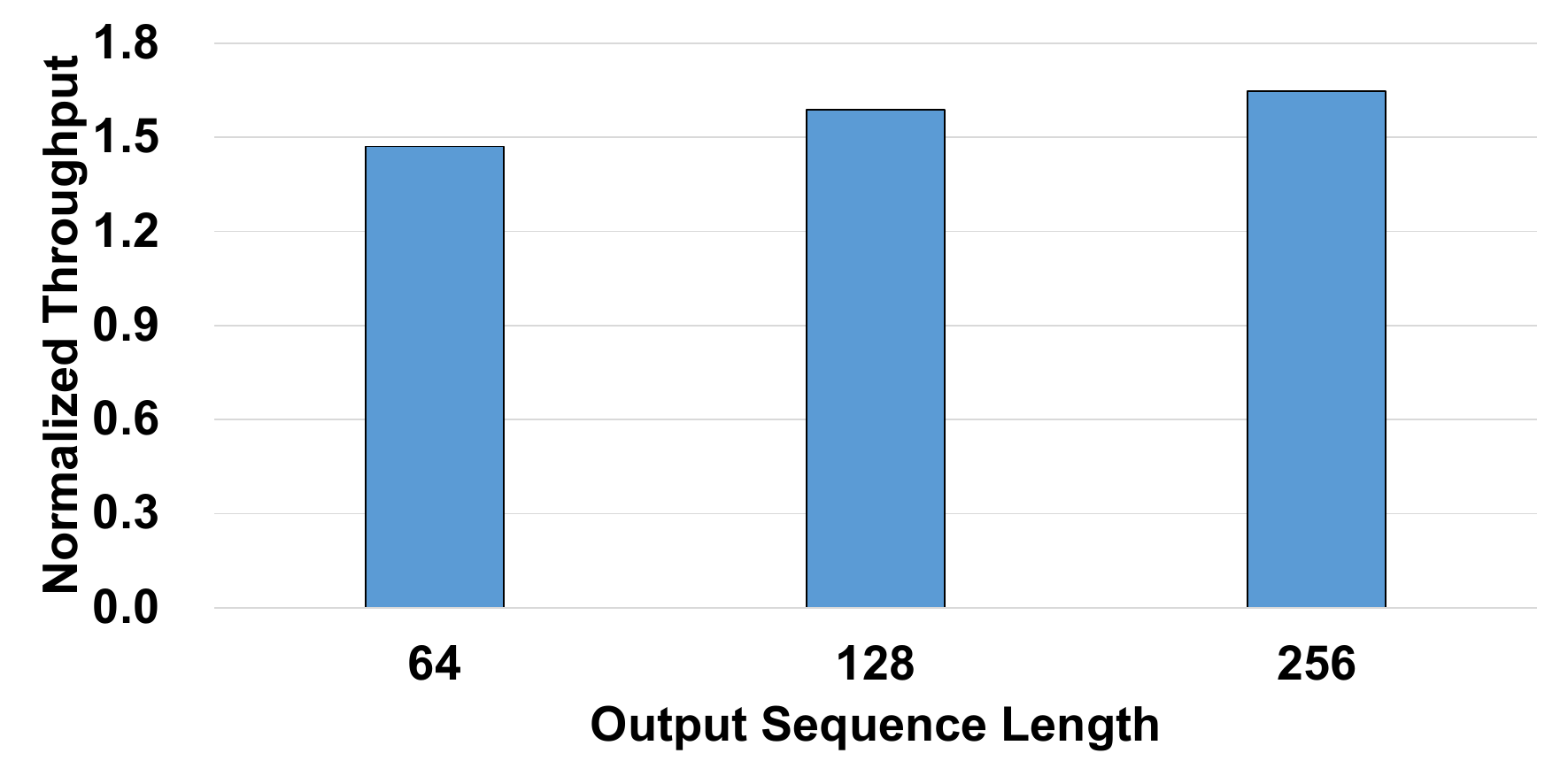}
\caption{Sensitivity to output length.}
\label{fig:sensitivity_length}
\end{figure}

Figure~\ref{fig:sensitivity_length} shows the effect of varying output length.
\textsc{MemSpec} achieves larger gains as generation becomes longer. Compared
with short outputs, longer generations provide more opportunity to exploit
intra-sequence variation and amortize the cost of draft adaptation. As a
result, runtime draft switching becomes increasingly beneficial as output
length grows.

This trend also indicates that the benefit of memory-aware scheduling becomes
more pronounced as generation complexity increases. Longer outputs typically
contain more phase transitions and greater intra-sequence heterogeneity,
making static draft selection increasingly suboptimal. In such settings,
proactively adapting the resident working set provides greater benefit than in
short, relatively homogeneous generations.

These results further support the central motivation of \textsc{MemSpec}:
dynamic draft adaptation is especially important for longer and more
heterogeneous generation workloads, and the advantage of \textsc{MemSpec}
scales with the opportunity for adaptation.

\paragraph*{Resident cache capacity.}
We additionally evaluate sensitivity to resident cache capacity \(K\), which
controls how many draft models can remain resident simultaneously. Because the
Jetson Orin Nano platform used in the main evaluation can hold at most two draft
models under the default configuration, we conduct this additional analysis on a
larger-memory Jetson AGX Orin platform.

\begin{figure}[t]
\centering
\includegraphics[width=1.0\linewidth]{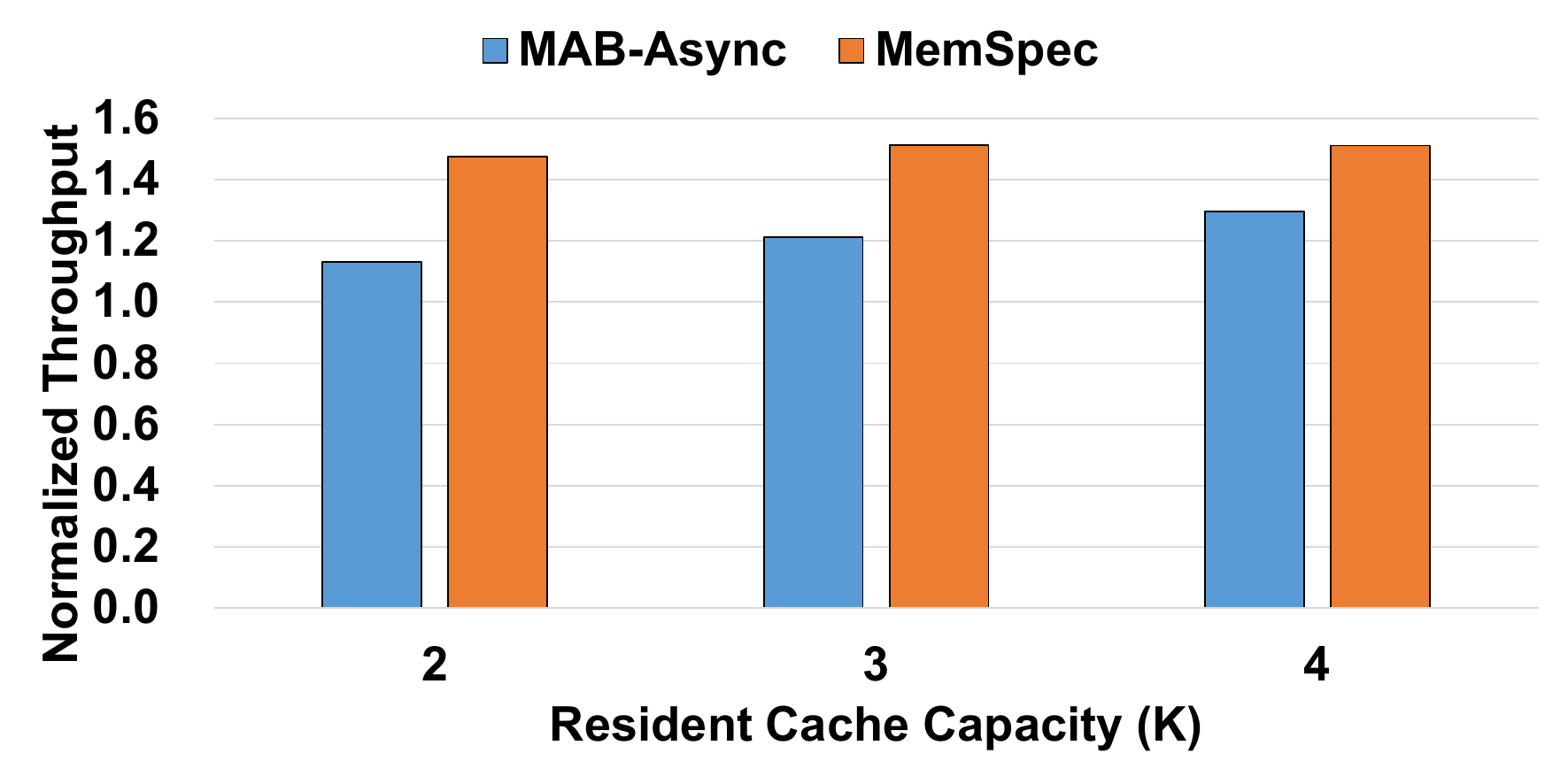}
\caption{Sensitivity to resident cache capacity \(K\) on Jetson AGX Orin.}
\label{fig:sensitivity_k}
\end{figure}

Figure~\ref{fig:sensitivity_k} shows that \textsc{MemSpec} already achieves
most of its performance benefit at relatively small cache capacities.
While \textit{MAB-Async} improves throughput from 1.13$\times$ to
1.29$\times$ as \(K\) increases from 2 to 4, \textsc{MemSpec} shows only
modest improvement (from 1.47$\times$ to 1.51$\times$). This behavior is
consistent with the predictor analysis in
Section~\ref{subsec:discussion}, which showed high top-2 recall. Since
high-utility drafts are already included in the resident working set in most
cases, increasing cache capacity further provides limited additional benefit
for \textsc{MemSpec}.

In contrast, \textit{MAB-Async} benefits more noticeably from larger resident
cache capacity. As \(K\) increases, reactive draft loading becomes less
frequent because the runtime experiences fewer non-resident draft selections.
Consequently, the performance gap between \textsc{MemSpec} and
\textit{MAB-Async} decreases slightly at larger \(K\). Nevertheless,
\textsc{MemSpec} consistently achieves higher throughput across all evaluated
cache capacities, demonstrating that prediction-guided residency management
remains beneficial even when more memory is available.

\subsection{Discussion}
\label{subsec:discussion}

\paragraph*{Predictor quality.}
\textsc{MemSpec} does not require perfectly accurate draft prediction to
improve throughput. The predictor is primarily used to identify promising
drafts for proactive residency management rather than to select the exact
optimal draft at every scheduling point.

To evaluate predictor quality, we compare three predictor input
configurations: prompt-only, recent-generated-tokens-only, and a combined
configuration using both prompt and recent generated tokens.

\begin{table}[t]
\centering
\caption{Predictor quality comparison.}
\label{tab:predictor_quality}

\setlength{\tabcolsep}{6pt}

\begin{tabular}{lcc}
\toprule
Input & Top-1 Acc. (\%) & Top-2 Recall (\%) \\
\midrule
Prompt-only & 31.9 & 55.8 \\
Recent-only & 57.1 & 79.3 \\
Prompt + Recent & 71.6 & 95.7 \\
\bottomrule
\end{tabular}
\end{table}

The combined predictor achieves the best ranking quality, improving top-1
accuracy to 71.6\% and achieving a top-2 recall of 95.7\%. This indicates
that the optimal draft is almost always included in the resident working set.
These results support the key design principle of \textsc{MemSpec}: adaptive
scheduling mainly requires sufficiently accurate ranking to maintain
high-utility drafts in the resident set, rather than perfect prediction.

\paragraph*{End-to-end latency considerations.}
Our main evaluation focuses on steady-state iterative decoding and therefore
excludes prompt processing and initial model loading. To assess startup
overhead, we additionally measure end-to-end latency including prompt
processing, target-model prefill, initial draft loading, and draft-model
preparation.

Although including startup costs reduces relative gains, \textsc{MemSpec}
still reduces end-to-end latency by 32.3\% on average over
\textit{General-Static} and by 24.9\% over \textit{MAB-Async}. These results
remain consistent with our main findings: the primary benefit of
\textsc{MemSpec} comes from reducing repeated draft loading and fallback
execution during iterative decoding.

\section{Related Work}

\subsection{Speculative Decoding and Adaptive Draft Selection}

Speculative decoding accelerates autoregressive LLM inference by using a
lightweight draft model to propose candidate tokens that are then verified by a
larger target
model~\cite{10.5555/3618408.3619203,chen2023acceleratinglargelanguagemodel}.
Prior work has improved this framework along multiple directions, including
specialized or distilled draft
models~\cite{zhou2024distillspec,yi-etal-2024-towards} and system optimizations
that reduce verification overhead or improve pipeline
efficiency~\cite{10.1145/3620666.3651335,pmlr-v262-mamou24a,zhang-etal-2025-draft,liu2025pearl}.

More recent studies observe that draft effectiveness varies across inputs
and even across generation stages within a single sequence. This has led to
adaptive speculative decoding methods that select among multiple candidate
drafts at runtime rather than relying on a single fixed
drafter~\cite{kim2025a,liu2026notabandit,hou2025banditspec}. In particular,
multi-drafter and bandit-based approaches improve acceptance through online
exploration and adaptation to observed runtime behavior.

However, these methods primarily optimize \emph{which} draft to select,
implicitly assuming that the chosen draft can be executed immediately. That
assumption is often invalid on memory-constrained edge devices, where only a
small number of drafts can remain resident and switching to a non-resident
draft incurs substantial loading overhead.

\textsc{MemSpec} differs from prior adaptive speculative decoding work by
treating draft selection and draft availability as a coupled systems problem.
Rather than focusing only on acceptance-rate improvement, it explicitly accounts
for residency and switching cost so that adaptive selection translates into
throughput gains under tight memory budgets.

\subsection{Adaptive Model Selection and Online Routing}

A broad line of research studies adaptive model selection in settings such as
cascaded inference~\cite{narasimhan2025faster,varshney-baral-2022-model,dekoninck2025a}
and model routing~\cite{chen2024frugalgpt,taylor2018adaptive,dekoninck2025a,pmlr-v162-chen22ad}.
These approaches dynamically select among candidate models based on input
difficulty, confidence estimates, or runtime feedback to balance accuracy and
efficiency. Some methods further employ online learning or bandit algorithms to
manage the exploration--exploitation trade-off.

\textsc{MemSpec} is related to this literature in that it also performs runtime
model selection, but the underlying systems constraints are fundamentally
different. Conventional routing formulations typically assume that all candidate
models are readily available for immediate execution, and thus optimize only the
selection decision itself.

In contrast, \textsc{MemSpec} operates in a setting where model
\emph{availability} is constrained by memory capacity and model loading
latency. Unlike conventional adaptive routing settings, speculative decoding
requires repeated draft decisions within a single generation trajectory,
making switching overhead a first-order cost. \textsc{MemSpec} therefore
extends adaptive model selection with residency-aware scheduling, separating
which drafts should be prepared from which resident draft should be executed
immediately.

\subsection{Memory-Constrained LLM Inference on Edge Devices}

Running LLMs on edge devices has motivated extensive work on reducing memory
footprint and managing limited memory resources. Existing approaches include
model compression (e.g., quantization, distillation) to shrink model
size~\cite{ko2025distillm,frantar2023optq,pmlr-v202-xiao23c,MLSYS2024_42a452cb}
and parameter offloading across GPU, CPU, and storage
tiers~\cite{NEURIPS2024_1d91d568,10.5555/3618408.3619696,10.1145/3694715.3695964,alizadeh-etal-2024-llm,10.1145/3656019.3676945}.
These techniques make large-model inference feasible on resource-limited
platforms.

Our work is complementary to these efforts but addresses a different challenge.
Prior memory-optimization methods mainly focus on optimizing the execution of a
\emph{single} large model by reducing its footprint or by moving its parameters
more efficiently across memory tiers.

By contrast, \textsc{MemSpec} targets adaptive speculative decoding with
\emph{multiple} candidate draft models, where the main challenge is not only
executing one model efficiently, but deciding which drafts should remain
resident, which should be prefetched, and when switching is worthwhile.

In this sense, prior edge inference techniques reduce the per-model memory
cost, while \textsc{MemSpec} addresses the runtime scheduling problem of
managing multiple drafts under a tight memory budget. These directions are
orthogonal and potentially complementary.

\section{Conclusion}

This paper presents \textsc{MemSpec}, a memory-aware runtime system for adaptive
draft scheduling in speculative decoding on memory-constrained edge platforms.
Through detailed characterization on a Jetson Orin Nano device, we show that
although generation heterogeneity creates substantial opportunities for
improving token acceptance, adaptive draft switching incurs significant model
loading overhead---often exceeding multiple decoding iterations---which limits
end-to-end throughput gains.

To address this challenge, \textsc{MemSpec} formulates adaptive speculative
decoding as a residency-constrained scheduling problem. By combining
prediction-guided draft ranking with memory-aware residency management,
\textsc{MemSpec} proactively aligns draft selection with runtime availability,
allowing decoding to proceed with the best currently resident draft while
asynchronously preparing better candidates in the background. Experimental
results show that \textsc{MemSpec} improves steady-state generation throughput
by $40.7\%$ on average over state-of-the-art adaptive baselines, while
achieving $95$--$97\%$ of an oracle dynamic upper bound under the same memory
constraints. These results demonstrate that the primary bottleneck in adaptive
speculative decoding on edge devices is not draft selection quality, but
ensuring timely availability of effective drafts.

\section*{Acknowledgements}

This research was supported in part by the National Research Foundation of
Korea (NRF) grant funded by the Korean government (MSIT) (No.
RS-2025-24535034), and in part by the Advanced GPU Utilization Support Program
funded by the Korean government (MSIT). Myeonggyun Han is the corresponding
author.

\bibliographystyle{ACM-Reference-Format}
\balance
\bibliography{./refs}

@article{10906629,
  author =        {Yi, Rongjie and Guo, Liwei and Wei, Shiyun and
                   Zhou, Ao and Wang, Shangguang and Xu, Mengwei},
  journal =       {IEEE Transactions on Mobile Computing},
  number =        {8},
  pages =         {7059-7073},
  title =         {{EdgeMoE}: Empowering Sparse Large Language Models on Mobile Devices},
  volume =        {24},
  year =          {2025},
  doi =           {10.1109/TMC.2025.3546466},
}

@inproceedings{10.1145/3649329.3655665,
  address =       {New York, NY, USA},
  author =        {Qin, Ruiyang and Xia, Jun and Jia, Zhenge and
                   Jiang, Meng and Abbasi, Ahmed and Zhou, Peipei and
                   Hu, Jingtong and Shi, Yiyu},
  booktitle =     {Proceedings of the 61st ACM/IEEE Design Automation Conference (DAC '24')},
  title =         {Enabling On-Device Large Language Model
                   Personalization with Self-Supervised Data Selection
                   and Synthesis},
  year =          {2024},
  doi =           {10.1145/3649329.3655665},
}

@inproceedings{10.5555/3618408.3619203,
  address =       {Honolulu, Hawaii, USA},
  author =        {Leviathan, Yaniv and Kalman, Matan and Matias, Yossi},
  booktitle =     {Proceedings of the International Conference on Machine Learning (ICML)},
  title =         {Fast Inference from Transformers via Speculative Decoding},
  year =          {2023},
}

@inproceedings{10.5555/3692070.3692273,
  address =       {Vienna, Austria},
  author =        {Cai, Tianle and Li, Yuhong and Geng, Zhengyang and
                   Peng, Hongwu and Lee, Jason D. and Chen, Deming and
                   Dao, Tri},
  booktitle =     {Proceedings of the International Conference on Machine Learning (ICML)},
  title =         {{MEDUSA}: Simple {LLM} Inference Acceleration Framework with Multiple Decoding Heads},
  year =          {2024},
}

@inproceedings{10.1145/3620666.3651335,
  address =       {New York, NY, USA},
  author =        {Miao, Xupeng and Oliaro, Gabriele and Zhang, Zhihao and
                   Cheng, Xinhao and Wang, Zeyu and Zhang, Zhengxin and
                   Wong, Rae Ying Yee and Zhu, Alan and Yang, Lijie and
                   Shi, Xiaoxiang and Shi, Chunan and Chen, Zhuoming and
                   Arfeen, Daiyaan and Abhyankar, Reyna and Jia, Zhihao},
  booktitle =     {Proceedings of the 29th ACM International Conference on Architectural Support for Programming Languages and Operating Systems, Volume 3 (ASPLOS '24')},
  pages =         {932--949},
  title =         {{SpecInfer}: Accelerating Large Language Model Serving with Tree-Based Speculative Inference and Verification},
  year =          {2024},
  doi =           {10.1145/3620666.3651335},
}

@inproceedings{yi-etal-2024-towards,
  address =       {Miami, Florida, USA},
  author =        {Yi, Euiin and Kim, Taehyeon and Jeung, Hongseok and
                   Chang, Du-Seong and Yun, Se-Young},
  booktitle =     {Proceedings of the Conference on Empirical Methods in Natural Language Processing (EMNLP)},
  editor =        {Al-Onaizan, Yaser and Bansal, Mohit and
                   Chen, Yun-Nung},
  month =         nov,
  pages =         {10789--10802},
  title =         {Towards Fast Multilingual {LLM} Inference: Speculative Decoding and Specialized Drafters},
  year =          {2024},
  doi =           {10.18653/v1/2024.emnlp-main.602},
}

@misc{kim2025a,
  author =        {Taehyeon Kim and Hojung Jung and Se-Young Yun},
  title =         {A Unified Framework for Speculative Decoding with Multiple Drafters as a Bandit},
  year =          {2025},
  url =           {https://openreview.net/forum?id=5haYLrlyGj},
}

@inproceedings{yan-etal-2025-decoding,
  address =       {Albuquerque, New Mexico},
  author =        {Yan, Minghao and Agarwal, Saurabh and
                   Venkataraman, Shivaram},
  booktitle =     {Proceedings of the Annual Conference of the Nations of the Americas Chapter of the Association for Computational Linguistics (NAACL)},
  editor =        {Chiruzzo, Luis and Ritter, Alan and Wang, Lu},
  month =         apr,
  pages =         {6460--6473},
  title =         {Decoding Speculative Decoding},
  year =          {2025},
  doi =           {10.18653/v1/2025.naacl-long.328},
}

@inproceedings{hou2025banditspec,
  address =       {Vancouver, Canada},
  author =        {Yunlong Hou and Fengzhuo Zhang and Cunxiao Du and
                   Xuan Zhang and Jiachun Pan and Tianyu Pang and
                   Chao Du and Vincent Y. F. Tan and Zhuoran Yang},
  booktitle =     {Proceedings of the International Conference on Machine Learning (ICML)},
  title =         {{BanditSpec}: Adaptive Speculative Decoding via Bandit Algorithms},
  volume =        {267},
  year =          {2025},
  url =           {https://openreview.net/forum?id=ghkWIlliZ8},
}

@inproceedings{liu2026notabandit,
  address =       {Rio de Janeiro, Brazil},
  author =        {Hongyi Liu and Jiaji Huang and Zhen Jia and
                   Youngsuk Park and Yu-Xiang Wang},
  booktitle =     {Proceedings of the International Conference on Learning Representations (ICLR)},
  title =         {Not-a-Bandit: Provably No-Regret Drafter Selection in
                   Speculative Decoding for {LLM}s},
  year =          {2026},
  url =           {https://openreview.net/forum?id=JMmljf895g},
}

@inproceedings{NEURIPS2024_1d91d568,
  address =       {Vancouver, Canada},
  author =        {Svirschevski, Ruslan and May, Avner and
                   Chen, Zhuoming and Chen, Beidi and Jia, Zhihao and
                   Ryabinin, Max},
  booktitle =     {Advances in Neural Information Processing Systems (NeurIPS)},
  editor =        {A. Globerson and L. Mackey and D. Belgrave and A. Fan and
                   U. Paquet and J. Tomczak and C. Zhang},
  pages =         {16342--16368},
  title =         {{SpecExec}: Massively Parallel Speculative Decoding for Interactive {LLM} Inference on Consumer Devices},
  volume =        {37},
  year =          {2024},
  url =           {https://proceedings.neurips.cc/paper_files/paper/2024/file/
                  1d91d5689e251d27993a3c2182dddcf7-Paper-Conference.pdf},
}

@misc{touvron2023llama2openfoundation,
  author =        {Hugo Touvron and Louis Martin and Kevin Stone and
                   Peter Albert and Amjad Almahairi and Yasmine Babaei and
                   Nikolay Bashlykov and Soumya Batra and
                   Prajjwal Bhargava and Shruti Bhosale and Dan Bikel and
                   Lukas Blecher and Cristian Canton Ferrer and
                   Moya Chen and Guillem Cucurull and David Esiobu and
                   Jude Fernandes and Jeremy Fu and Wenyin Fu and
                   Brian Fuller and Cynthia Gao and Vedanuj Goswami and
                   Naman Goyal and Anthony Hartshorn and Saghar Hosseini and
                   Rui Hou and Hakan Inan and Marcin Kardas and
                   Viktor Kerkez and Madian Khabsa and Isabel Kloumann and
                   Artem Korenev and Punit Singh Koura and
                   Marie-Anne Lachaux and Thibaut Lavril and Jenya Lee and
                   Diana Liskovich and Yinghai Lu and Yuning Mao and
                   Xavier Martinet and Todor Mihaylov and Pushkar Mishra and
                   Igor Molybog and Yixin Nie and Andrew Poulton and
                   Jeremy Reizenstein and Rashi Rungta and Kalyan Saladi and
                   Alan Schelten and Ruan Silva and Eric Michael Smith and
                   Ranjan Subramanian and Xiaoqing Ellen Tan and
                   Binh Tang and Ross Taylor and Adina Williams and
                   Jian Xiang Kuan and Puxin Xu and Zheng Yan and
                   Iliyan Zarov and Yuchen Zhang and Angela Fan and
                   Melanie Kambadur and Sharan Narang and
                   Aurelien Rodriguez and Robert Stojnic and
                   Sergey Edunov and Thomas Scialom},
  title =         {{Llama 2}: Open Foundation and Fine-Tuned Chat Models},
  year =          {2023},
  eprint =        {2307.09288},
  archivePrefix = {arXiv},
}

@misc{qwen2025qwen25technicalreport,
  author =        {Qwen Team and An Yang and Baosong Yang and
                   Beichen Zhang and Binyuan Hui and Bo Zheng and
                   Bowen Yu and Chengyuan Li and Dayiheng Liu and
                   Fei Huang and Haoran Wei and Huan Lin and Jian Yang and
                   Jianhong Tu and Jianwei Zhang and Jianxin Yang and
                   Jiaxi Yang and Jingren Zhou and Junyang Lin and
                   Kai Dang and Keming Lu and Keqin Bao and Kexin Yang and
                   Le Yu and Mei Li and Mingfeng Xue and Pei Zhang and
                   Qin Zhu and Rui Men and Runji Lin and Tianhao Li and
                   Tianyi Tang and Tingyu Xia and Xingzhang Ren and
                   Xuancheng Ren and Yang Fan and Yang Su and
                   Yichang Zhang and Yu Wan and Yuqiong Liu and Zeyu Cui and
                   Zhenru Zhang and Zihan Qiu},
  title =         {{Qwen2.5} Technical Report},
  year =          {2025},
  eprint =        {2412.15115},
  archivePrefix = {arXiv},
}

@misc{cobbe2021trainingverifierssolvemath,
  author =        {Karl Cobbe and Vineet Kosaraju and Mohammad Bavarian and
                   Mark Chen and Heewoo Jun and Lukasz Kaiser and
                   Matthias Plappert and Jerry Tworek and Jacob Hilton and
                   Reiichiro Nakano and Christopher Hesse and
                   John Schulman},
  title =         {Training Verifiers to Solve Math Word Problems},
  year =          {2021},
  eprint =        {2110.14168},
  archivePrefix = {arXiv},
}

@inproceedings{hendrycks2021measuring,
  address =       {New Orleans, LA, USA},
  author =        {Hendrycks, Dan and Burns, Collin and Kadavath, Saurav and
                   Arora, Akul and Basart, Steven and Tang, Eric and
                   Song, Dawn and Steinhardt, Jacob},
  booktitle =     {Proceedings of the Neural Information Processing
                   Systems Track on Datasets and Benchmarks},
  title =         {Measuring Mathematical Problem Solving With the
                   {MATH} Dataset},
  year =          {2021},
  url =           {https://openreview.net/forum?id=7Bywt2mQsCe},
}

@misc{chen2021evaluating,
  author =        {Mark Chen and Jerry Tworek and Heewoo Jun and
                   Qiming Yuan and Henrique Ponde de Oliveira Pinto and
                   Jared Kaplan and Harri Edwards and Yuri Burda and
                   Nicholas Joseph and Greg Brockman and Alex Ray and
                   Raul Puri and Gretchen Krueger and Michael Petrov and
                   Heidy Khlaaf and Girish Sastry and Pamela Mishkin and
                   Brooke Chan and Scott Gray and Nick Ryder and
                   Mikhail Pavlov and Alethea Power and Lukasz Kaiser and
                   Mohammad Bavarian and Clemens Winter and
                   Philippe Tillet and Felipe Petroski Such and
                   Dave Cummings and Matthias Plappert and
                   Fotios Chantzis and Elizabeth Barnes and
                   Ariel Herbert-Voss and William Hebgen Guss and
                   Alex Nichol and Alex Paino and Nikolas Tezak and
                   Jie Tang and Igor Babuschkin and Suchir Balaji and
                   Shantanu Jain and William Saunders and
                   Christopher Hesse and Andrew N. Carr and Jan Leike and
                   Josh Achiam and Vedant Misra and Evan Morikawa and
                   Alec Radford and Matthew Knight and Miles Brundage and
                   Mira Murati and Katie Mayer and Peter Welinder and
                   Bob McGrew and Dario Amodei and Sam McCandlish and
                   Ilya Sutskever and Wojciech Zaremba},
  title =         {Evaluating Large Language Models Trained on Code},
  year =          {2021},
  eprint =        {2107.03374},
  archivePrefix = {arXiv},
}

@misc{austin2021programsynthesislargelanguage,
  author =        {Jacob Austin and Augustus Odena and Maxwell Nye and
                   Maarten Bosma and Henryk Michalewski and David Dohan and
                   Ellen Jiang and Carrie Cai and Michael Terry and
                   Quoc Le and Charles Sutton},
  title =         {Program Synthesis with Large Language Models},
  year =          {2021},
  eprint =        {2108.07732},
  archivePrefix = {arXiv},
}

@inproceedings{chalkidis-etal-2022-lexglue,
  address =       {Dublin, Ireland},
  author =        {Chalkidis, Ilias and Jana, Abhik and Hartung, Dirk and
                   Bommarito, Michael and Androutsopoulos, Ion and
                   Katz, Daniel and Aletras, Nikolaos},
  booktitle =     {Proceedings of the Annual Meeting of the Association for Computational Linguistics (ACL)},
  editor =        {Muresan, Smaranda and Nakov, Preslav and
                   Villavicencio, Aline},
  month =         may,
  pages =         {4310--4330},
  title =         {{L}ex{GLUE}: A Benchmark Dataset for Legal Language
                   Understanding in {E}nglish},
  year =          {2022},
  doi =           {10.18653/v1/2022.acl-long.297},
}

@article{app11146421,
  author =        {Jin, Di and Pan, Eileen and Oufattole, Nassim and
                   Weng, Wei-Hung and Fang, Hanyi and Szolovits, Peter},
  journal =       {Applied Sciences},
  number =        {14},
  title =         {What Disease Does This Patient Have? A Large-Scale
                   Open Domain Question Answering Dataset from Medical
                   Exams},
  volume =        {11},
  year =          {2021},
  doi =           {10.3390/app11146421},
  url =           {https://www.mdpi.com/2076-3417/11/14/6421},
}

@inproceedings{pmlr-v174-pal22a,
  address =       {415 Main Street, Cambridge, MA USA},
  author =        {Pal, Ankit and Umapathi, Logesh Kumar and
                   Sankarasubbu, Malaikannan},
  booktitle =     {Proceedings of the Conference on Health, Inference,
                   and Learning},
  editor =        {Flores, Gerardo and Chen, George H and Pollard, Tom and
                   Ho, Joyce C and Naumann, Tristan},
  month =         {07--08 Apr},
  pages =         {248--260},
  title =         {{MedMCQA}: A Large-Scale Multi-Subject Multi-Choice Dataset for Medical Domain Question Answering},
  volume =        {174},
  year =          {2022},
  url =           {https://proceedings.mlr.press/v174/pal22a.html},
}

@misc{alpaca,
  author =        {Rohan Taori and Ishaan Gulrajani and Tianyi Zhang and
                   Yann Dubois and Xuechen Li and Carlos Guestrin and
                   Percy Liang and Tatsunori B. Hashimoto},
  howpublished =  {\url{https://github.com/tatsu-lab/stanford_alpaca}},
  title =         {Stanford {Alpaca}: An Instruction-Following {LLaMA} Model},
  year =          {2023},
}

@misc{jain2025livecodebench,
  author =        {Naman Jain and King Han and Alex Gu and Wen-Ding Li and
                   Fanjia Yan and Tianjun Zhang and Sida Wang and
                   Armando Solar-Lezama and Koushik Sen and Ion Stoica},
  title =         {{LiveCodeBench}: Holistic and Contamination Free Evaluation of Large Language Models for Code},
  year =          {2024},
  eprint =        {2403.07974},
  archivePrefix = {arXiv},
}

@misc{gao2024omnimathuniversalolympiadlevel,
  author =        {Bofei Gao and Feifan Song and Zhe Yang and Zefan Cai and
                   Yibo Miao and Qingxiu Dong and Lei Li and Chenghao Ma and
                   Liang Chen and Runxin Xu and Zhengyang Tang and
                   Benyou Wang and Daoguang Zan and Shanghaoran Quan and
                   Ge Zhang and Lei Sha and Yichang Zhang and
                   Xuancheng Ren and Tianyu Liu and Baobao Chang},
  title =         {{Omni-MATH}: A Universal Olympiad Level Mathematic Benchmark for Large Language Models},
  year =          {2024},
  eprint =        {2410.07985},
  archivePrefix = {arXiv},
}

@misc{chen2023acceleratinglargelanguagemodel,
  author =        {Charlie Chen and Sebastian Borgeaud and
                   Geoffrey Irving and Jean-Baptiste Lespiau and
                   Laurent Sifre and John Jumper},
  title =         {Accelerating Large Language Model Decoding with
                   Speculative Sampling},
  year =          {2023},
  eprint =        {2302.01318},
  archivePrefix = {arXiv},
}

@inproceedings{zhou2024distillspec,
  address =       {Vienna, Austria},
  author =        {Yongchao Zhou and Kaifeng Lyu and Ankit Singh Rawat and
                   Aditya Krishna Menon and Afshin Rostamizadeh and
                   Sanjiv Kumar and Jean-Fran{\c{c}}ois Kagy and
                   Rishabh Agarwal},
  booktitle =     {Proceedings of the International Conference on Learning Representations (ICLR)},
  title =         {{DistillSpec}: Improving Speculative Decoding via Knowledge Distillation},
  year =          {2024},
  url =           {https://openreview.net/forum?id=rsY6J3ZaTF},
}

@inproceedings{pmlr-v262-mamou24a,
  address =       {Vancouver, Canada},
  author =        {Mamou, Jonathan and Pereg, Oren and Korat, Daniel and
                   Berchansky, Moshe and Timor, Nadav and
                   Wasserblat, Moshe and Schwartz, Roy},
  booktitle =     {Proceedings of The 4th NeurIPS Efficient Natural
                   Language and Speech Processing Workshop},
  editor =        {Rezagholizadeh, Mehdi and Passban, Peyman and
                   Samiee, Soheila and Partovi Nia, Vahid and Cheng, Yu and
                   Deng, Yue and Liu, Qun and Chen, Boxing},
  month =         {14 Dec},
  pages =         {456--467},
  title =         {Dynamic Speculation Lookahead Accelerates Speculative
                   Decoding of Large Language Models},
  volume =        {262},
  year =          {2024},
  url =           {https://proceedings.mlr.press/v262/mamou24a.html},
}

@inproceedings{zhang-etal-2025-draft,
  address =       {Suzhou, China},
  author =        {Zhang, Ziyin and Xu, Jiahao and Liang, Tian and
                   Chen, Xingyu and He, Zhiwei and Wang, Rui and
                   Tu, Zhaopeng},
  booktitle =     {Proceedings of the Conference on Empirical Methods in Natural Language Processing (EMNLP)},
  editor =        {Christodoulopoulos, Christos and Chakraborty, Tanmoy and
                   Rose, Carolyn and Peng, Violet},
  month =         nov,
  pages =         {16696--16708},
  title =         {Draft Model Knows When to Stop: Self-Verification
                   Speculative Decoding for Long-Form Generation},
  year =          {2025},
  doi =           {10.18653/v1/2025.emnlp-main.844},
}

@inproceedings{liu2025pearl,
  address =       {Singapore},
  author =        {Tianyu Liu and Yun Li and Qitan Lv and Kai Liu and
                   Jianchen Zhu and Winston Hu and Xiao Sun},
  booktitle =     {Proceedings of the International Conference on Learning Representations (ICLR)},
  title =         {{PEARL}: Parallel Speculative Decoding with Adaptive Draft Length},
  year =          {2025},
  url =           {https://openreview.net/forum?id=QOXrVMiHGK},
}

@article{chen2024frugalgpt,
  author =        {Lingjiao Chen and Matei Zaharia and James Zou},
  journal =       {Transactions on Machine Learning Research},
  title =         {{FrugalGPT}: How to Use Large Language Models While Reducing Cost and Improving Performance},
  volume =        {2024},
  year =          {2024},
  url =           {https://openreview.net/forum?id=cSimKw5p6R},
}

@inproceedings{taylor2018adaptive,
	author = {Ben Taylor and Vicent Sanz Marco and Willy Wolff and Yehia Elkhatib and Zheng Wang},
  title = {Adaptive Deep Learning Model Selection on Embedded Systems},
  year = {2018},
  booktitle = {Proceedings of the 19th ACM SIGPLAN/SIGBED International Conference on Languages, Compilers, and Tools for Embedded Systems (LCTES '18)},
  location = {Philadelphia, PA, USA},
  pages = {31--43},
  doi = {10.1145/3211332.3211336}
}

@inproceedings{dekoninck2025a,
  address =       {Vancouver, Canada},
  author =        {Jasper Dekoninck and Maximilian Baader and
                   Martin Vechev},
  booktitle =     {Proceedings of the International Conference on Machine Learning (ICML)},
  title =         {A Unified Approach to Routing and Cascading for
                   {LLM}s},
  volume =        {267},
  year =          {2025},
  url =           {https://openreview.net/forum?id=AAl89VNNy1},
}

@inproceedings{pmlr-v162-chen22ad,
  address =       {Baltimore, Maryland, USA},
  author =        {Chen, Lingjiao and Zaharia, Matei and Zou, James},
  booktitle =     {Proceedings of the 39th International Conference on
                   Machine Learning},
  editor =        {Chaudhuri, Kamalika and Jegelka, Stefanie and
                   Song, Le and Szepesvari, Csaba and Niu, Gang and
                   Sabato, Sivan},
  month =         {17--23 Jul},
  pages =         {3716--3746},
  title =         {Efficient Online {ML} {API} Selection for Multi-Label
                   Classification Tasks},
  volume =        {162},
  year =          {2022},
  url =           {https://proceedings.mlr.press/v162/chen22ad.html},
}

@inproceedings{varshney-baral-2022-model,
  address =       {Abu Dhabi, United Arab Emirates},
  author =        {Varshney, Neeraj and Baral, Chitta},
  booktitle =     {Proceedings of the Conference on Empirical Methods in Natural Language Processing (EMNLP)},
  editor =        {Goldberg, Yoav and Kozareva, Zornitsa and Zhang, Yue},
  month =         dec,
  pages =         {11007--11021},
  title =         {Model Cascading: Towards Jointly Improving Efficiency
                   and Accuracy of {NLP} Systems},
  year =          {2022},
  doi =           {10.18653/v1/2022.emnlp-main.756},
}

@inproceedings{ko2025distillm,
  address =       {Vancouver, Canada},
  author =        {Jongwoo Ko and Tianyi Chen and Sungnyun Kim and
                   Tianyu Ding and Luming Liang and Ilya Zharkov and
                   Se-Young Yun},
  booktitle =     {Proceedings of the International Conference on Machine Learning (ICML)},
  title =         {{DistiLLM}-2: A Contrastive Approach Boosts the Distillation of {LLM}s},
  volume =        {267},
  year =          {2025},
  url =           {https://openreview.net/forum?id=rc65N9xIrY},
}

@inproceedings{frantar2023optq,
  address =       {Kigali, Rwanda},
  author =        {Elias Frantar and Saleh Ashkboos and Torsten Hoefler and
                   Dan Alistarh},
  booktitle =     {Proceedings of the International Conference on Learning Representations (ICLR)},
  title =         {{OPTQ}: Accurate Quantization for Generative Pre-trained Transformers},
  year =          {2023},
  url =           {https://openreview.net/forum?id=tcbBPnfwxS},
}

@inproceedings{pmlr-v202-xiao23c,
  address =       {Honolulu, Hawaii, USA},
  author =        {Xiao, Guangxuan and Lin, Ji and Seznec, Mickael and
                   Wu, Hao and Demouth, Julien and Han, Song},
  booktitle =     {Proceedings of the International Conference on Machine Learning (ICML)},
  editor =        {Krause, Andreas and Brunskill, Emma and
                   Cho, Kyunghyun and Engelhardt, Barbara and
                   Sabato, Sivan and Scarlett, Jonathan},
  month =         {23--29 Jul},
  pages =         {38087--38099},
  title =         {{S}mooth{Q}uant: Accurate and Efficient Post-Training
                   Quantization for Large Language Models},
  volume =        {202},
  year =          {2023},
  url =           {https://proceedings.mlr.press/v202/xiao23c.html},
}

@inproceedings{MLSYS2024_42a452cb,
  address =       {Santa Clara, CA, USA},
  author =        {Lin, Ji and Tang, Jiaming and Tang, Haotian and
                   Yang, Shang and Chen, Wei-Ming and Wang, Wei-Chen and
                   Xiao, Guangxuan and Dang, Xingyu and Gan, Chuang and
                   Han, Song},
  booktitle =     {Proceedings of Machine Learning and Systems (MLSys)},
  editor =        {P. Gibbons and G. Pekhimenko and C. De Sa},
  pages =         {87--100},
  title =         {{AWQ}: Activation-Aware Weight Quantization for On-Device {LLM} Compression and Acceleration},
  volume =        {6},
  year =          {2024},
  url =           {https://proceedings.mlsys.org/paper_files/paper/2024/file/
                  42a452cbafa9dd64e9ba4aa95cc1ef21-Paper-Conference.pdf},
}

@inproceedings{10.5555/3618408.3619696,
  author =        {Sheng, Ying and Zheng, Lianmin and Yuan, Binhang and
                   Li, Zhuohan and Ryabinin, Max and Chen, Beidi and
                   Liang, Percy and R\'{e}, Christopher and Stoica, Ion and
                   Zhang, Ce},
  booktitle =     {Proceedings of the International Conference on Machine Learning (ICML)},
  title =         {{FlexGen}: High-Throughput Generative Inference of Large Language Models with a Single {GPU}},
  year =          {2023},
}

@inproceedings{10.1145/3694715.3695964,
  address =       {New York, NY, USA},
  author =        {Song, Yixin and Mi, Zeyu and Xie, Haotong and
                   Chen, Haibo},
  booktitle =     {Proceedings of the ACM Symposium on Operating Systems Principles (SOSP)},
  pages =         {590--606},
  title =         {{PowerInfer}: Fast Large Language Model Serving with a Consumer-Grade {GPU}},
  year =          {2024},
  doi =           {10.1145/3694715.3695964},
}

@inproceedings{alizadeh-etal-2024-llm,
  address =       {Bangkok, Thailand},
  author =        {Alizadeh, Keivan and Mirzadeh, Seyed Iman and
                   Belenko, Dmitry and Khatamifard, S. and Cho, Minsik and
                   Del Mundo, Carlo C and Rastegari, Mohammad and
                   Farajtabar, Mehrdad},
  booktitle =     {Proceedings of the Annual Meeting of the Association for Computational Linguistics (ACL)},
  editor =        {Ku, Lun-Wei and Martins, Andre and Srikumar, Vivek},
  month =         aug,
  pages =         {12562--12584},
  title =         {{LLM} in a flash: Efficient Large Language Model
                   Inference with Limited Memory},
  year =          {2024},
  doi =           {10.18653/v1/2024.acl-long.678},
}

@inproceedings{10.1145/3656019.3676945,
  address =       {New York, NY, USA},
  author =        {Kim, Sowoong and Sim, Eunyeong and Shin, Youngsam and
                   Cho, YeonGon and Baek, Woongki},
  booktitle =     {Proceedings of the 2024 International Conference on Parallel Architectures and Compilation Techniques (PACT '24')},
  pages =         {78--90},
  title =         {Activation Sequence Caching: High-Throughput and
                   Memory-Efficient Generative Inference with a Single
                   GPU},
  year =          {2024},
  doi =           {10.1145/3656019.3676945},
}

@inproceedings{narasimhan2025faster,
  author = {Harikrishna Narasimhan and Wittawat Jitkrittum and Ankit Singh Rawat
    and Seungyeon Kim and Neha Gupta and Aditya Krishna Menon and Sanjiv Kumar},
  title = {Faster Cascades via Speculative Decoding},
  booktitle = {Proceedings of the International Conference on Learning Representations (ICLR)},
  year = {2025},
  location = {Singapore},
  url = {https://openreview.net/forum?id=vo9t20wsmd}
}

\end{document}